\documentclass[cernpreprint,USenglish]{na61doc}
\usepackage{amsmath}
\usepackage{url}

\usepackage{colortbl}
\definecolor{darkred}{rgb}{0.5,0,0}
\definecolor{darkblue}{rgb}{0,0,0.5}
\definecolor{firebrick}{rgb}{0.75,0.125,0.125}
\definecolor{darkgreen}{rgb}{0,0.5,0}

\usepackage[colorlinks=true,linkcolor=firebrick,citecolor=darkgreen,urlcolor=darkblue]{hyperref}

\usepackage{xspace}
\usepackage{enumerate}
\usepackage{lineno}

\newcommand{\detadphi}{$\Delta\eta\Delta\phi$}

\PreprintIdNumber{CERN-EP-2020-103}
\ShineJournal{EPJ C}

\ShineTitle{Two-particle correlations in azimuthal angle and pseudorapidity in central $^7$Be+$^9$Be collisions at the CERN Super Proton Synchrotron}

\ShineAbstract{
A measurement of charged hadron pair correlations in two-dimensional \detadphi~ space is presented.
The analysis is based on total 30 million central Be+Be collisions observed in the \NASixtyOne detector at the CERN SPS for incident beam momenta of 19$A$, 30$A$, 40$A$, 75$A$, and 150$A$~\GeVc.
Measurements were carried out for unlike-sign and like-sign charge hadron pairs independently.
The $C(\Delta\eta,\Delta\phi)$ correlation functions were compared with results from a similar analysis on p+p interactions at similar beam momenta per nucleon.
General trends of the back-to-back correlations are similar in central Be+Be collisions and p+p interactions, but are suppressed in magnitude due to the increased combinatorial background.
Predictions from the \Epos and UrQMD models are compared to the measurements.
Evolution of an enhancement around $(\Delta\eta,\Delta\phi) = (0,0)$ with incident energy is observed in central Be+Be collisions.
It is not predicted by both models and almost non-existing in proton-proton collisions at the same momentum per nucleon.
}

\newcommand{\eV}{\ensuremath{\mbox{e\kern-0.1em V}}\xspace}
\newcommand{\GeV}{\ensuremath{\mbox{Ge\kern-0.1em V}}\xspace}
\newcommand{\MeV}{\ensuremath{\mbox{Me\kern-0.1em V}}\xspace}
\newcommand{\GeVc}{\ensuremath{\mbox{Ge\kern-0.1em V}\!/\!c}\xspace}
\newcommand{\GeVcc}{\ensuremath{\mbox{Ge\kern-0.1em V}\!/\!c^2}\xspace}
\newcommand{\AGeV}{\ensuremath{A\,\mbox{Ge\kern-0.1em V}}\xspace}
\newcommand{\AGeVc}{\ensuremath{A\,\mbox{Ge\kern-0.1em V}\!/\!c}\xspace}
\newcommand{\MeVc}{\ensuremath{\mbox{Me\kern-0.1em V}/c}\xspace}

\newcommand{\dd}{\ensuremath{{\text{d}}}\xspace}
\newcommand{\dedx}{\ensuremath{\dd E\!/\!\dd x}\xspace}

\newcommand{\Geant}{{\scshape Geant}\xspace}

\newcommand{\Epos}{{\scshape Epos}\xspace}

\newcommand{\CernVM}{\textsc{Cern\-\kern-0.05emVM}\xspace}

\begin{document}

\maketitle

\section{Introduction and motivation}

This paper presents experimental results on two-particle correlations in pseudorapidity and azimuthal angle of charged particles produced in central Be+Be collisions at 19$A$, 30$A$, 40$A$, 75$A$, and 150\AGeVc. The measurements were performed by the multi-purpose \NASixtyOne~\cite{Abgrall:2014fa} experiment at the CERN Super Proton Synchrotron (SPS). They are part of the strong interactions programme devoted to the study of the properties of the onset of deconfinement and search for the critical point of strongly interacting matter. Within this program a two-dimensional scan in collision energy and size of colliding nuclei recorded data on p+p, Be+Be, Ar+Sc, Xe+La, and Pb+Pb collisions and was completed in 2018. The expected signal of a critical point is a non-monotonic dependence of various fluctuation measures in such a scan; for a recent review see Ref.~\cite{Gazdzicki:2015ska}.

Apart from looking for critical point (CP) and quark-gluon plasma (QGP) signatures, it is of interest to study specific physical phenomena that happen during and after the collision. The two-particle correlation analysis in pseudorapidity ($\eta$) and azimuthal angle ($\phi$) allows to disentangle different correlation sources which may be directly connected with phenomena like jets, collective flow, resonance decays, quantum statistics effects, conservation laws, etc.

Measurements of two-particle correlations in pseudorapidity and azimuthal angle were first published by the ACM collaboration at the Intersecting Storage Rings (ISR)~\cite{Eggert:1974ek}. Two- and three-body decays of resonances ($\eta$, $\rho^0$, $\omega$) were found to provide the dominant contributions. Two structures were observed: an enhancement near $\Delta\phi = \pi$ (away-side) explained by the two-body decay scenario and another enhancement at $\Delta\phi \approx 0$ together with an azimuthal ridge (centered at $\Delta\eta \approx 0$) consistent with three-body decays.\footnote{$\Delta\eta$ and $\Delta\phi$ definitions are in Eq.~\ref{eq:deta_dphi}.} These features were confirmed at the higher collision energies of Relativistic Heavy Ion Collider (RHIC) by the PHOBOS~\cite{Alver:2007wy} collaboration. 

At RHIC and the Large Hadron Collider (LHC) parton scattering processes become important. In addition to high transverse momentum jets, studies of $\Delta\eta\Delta\phi$ correlations in p+p interactions as well as in collisions of heavy nuclei ~\cite{Porter:2005rc, Porter:2004jt, Abelev:2014mva, Alver:2008aa} found prominent structures explained as arising from the production of minijets, creating a large correlation peak at small opening angles $(\Delta\eta,\Delta\phi) \approx (0,0)$ and a broad structure along $\Delta\eta$ at $\Delta\phi \approx \pi$ (also referred to as away-side ridge). 

A study of two-particle correlations was already performed by \NASixtyOne in inelastic p+p interactions at SPS energies and reported in Ref.~\cite{Aduszkiewicz:2016mww}. The results show structures connected most probably to resonance decays, momentum conservation, and Bose-Einstein correlations. No clear sign of jet-like structure was observed (a more detailed search for jet-like structures was performed in Ref.~\cite{BMthesis}). 

This paper reports \NASixtyOne results from the next step in size of the collision system of two-particle correlations in $\Delta\eta$ and $\Delta\phi$ for the 5\% most central $^7$Be+$^9$Be collisions. The data were recorded in 2011, 2012 and 2013 using a secondary $^7$Be beam produced by fragmentation of the primary Pb beam from the CERN SPS~\cite{Abgrall:7Bebeam}.
The $^7$Be+$^9$Be collisions play a special role in the \NASixtyOne scan programme. The collision system composed of a $^7$Be and a $^9$Be nucleus has eight protons and eight neutrons, and thus is isospin symmetric. 
Within the \NASixtyOne scan programme the $^7$Be+$^9$Be collisions serve as the lowest mass isospin symmetric reference needed to study collisions of medium and large mass nuclei. This is of particular importance when data on proton-proton, neutron-proton and neutron-neutron are not available to construct the nucleon-nucleon reference~\cite{Gazdzicki:1991ih}. Finally, the latest RHIC and LHC results suggest that collective effects may also be developed in small (p+Pb, d+Au) or high-multiplicity p+p systems (Refs.~\cite{Nagle:2018nvi,Mohapatra:2018dwe,Bhalerao:2020ulk}).

Study of energy evolution of the near-side $\Delta\eta,\Delta\phi$ correlation in Be+Be is also  of interest from the point of view of possible formation and decays of small QGP hot-spots (Refs.~\cite{Lindenbaum:2003ma,Lindenbaum:2008sx,VanHove:1984zy}) because the products of the first stage of the interaction will undergo less scattering in surrounding matter than in the case of  heavy nucleus-nucleus reactions.

In this paper the pseudorapidity variable $\eta$ is calculated as $\eta = -\ln(\tan(\Theta/2))$, where $\tan(\Theta) = p_T/p_L$ with $p_T$ the transverse ($x,y$) and
$p_L$ the longitudinal ($z$) component of the particle momentum in the collision centre-of-mass system. The pion mass was assumed for all particles in the Lorentz transformation of $p_L$ measured in the laboratory system to the centre-of-mass system.
The azimuthal angle $\phi$ is the angle between the transverse momentum vector and the horizontal ($x$) axis.

\section{Two-particle correlations in pseudorapidity and azimuthal angle}\label{sec:detadphi}

Correlations studied in this paper were calculated as a function of the difference in pseudorapidity
($\eta$) and azimuthal angle ($\phi$) between two particles produced in the same event:
\begin{equation}
  \label{eq:deta_dphi}
  \Delta\eta = |{\eta}_1 - {\eta}_2|,
    \hspace{2cm}
    \Delta\phi = |{\phi}_1 - {\phi}_2|.
\end{equation}

The correlation function $C(\Delta\eta,\Delta\phi)$ is defined and calculated as:
\begin{equation}
  \label{eq:correlations}
  C(\Delta\eta,\Delta\phi)=
  \frac{N_{\text{mixed}}^{\text{pairs}}}{N_{\text{data}}^{\text{pairs}}}
  \frac{D(\Delta\eta,\Delta\phi)}{M(\Delta\eta,\Delta\phi)},
\end{equation}
where
  \begin{equation*}
  D(\Delta\eta,\Delta\phi)=\frac{d^2N_{\text{data}}}{d \Delta \eta d
    \Delta \phi}, \hspace{0.5cm} 
  M(\Delta\eta,\Delta\phi)=\frac{d^2N_{\text{mixed}}}{d \Delta \eta d \Delta
    \phi}
\label{eq:D_and_M}
\end{equation*}
are the distributions of particle pairs from the same (data) and from different (mixed) events, respectively. Distributions $D(\Delta\eta,\Delta\phi)$ and $M(\Delta\eta,\Delta\phi)$ were obtained by accumulating the number of pairs in intervals of $\Delta\eta$ and $\Delta\phi$. For the calculation of $C(\Delta\eta,\Delta\phi)$ both distributions were normalised to the number of pairs ($N^{\text{pairs}}_{\text{data}}$, $N^{\text{pairs}}_{\text{mixed}}$) in the given distribution.

The uncorrelated background was constructed by mixing particles from different data events with two main constraints: (a) the multiplicity distribution of mixed events had to be exactly the same as the original data event; (b) mixed events could not contain two particles from the same data event. The same method of mixing events was used in Ref.~\cite{Aduszkiewicz:2016mww}.

As stated in Eq.~\ref{eq:deta_dphi}, the $\Delta\eta$ and $\Delta\phi$ values will get only positive values. Hence, the measurements were restricted to $0 \leq \Delta\eta \leq 3$ and $0 \leq \Delta\phi < \pi$. However, in order to better show the correlation structure, the results were mirrored along $\Delta\eta = 0$ and, assuming the periodicity in the range of $2\pi$, also along $\Delta\phi =0$ and $\Delta\phi = \pi$. Then, the distributions were shifted to the range $-\frac{\pi}{2} < \Delta\phi < \frac{3 \pi}{2}$. These modifications were done to better demonstrate the most interesting correlation structures as well as to make them easily comparable with results obtained by other experiments (e.g. from Refs.~\cite{CMS:2012qk,Khachatryan:2015lva}).

In this paper, the correlation function $C(\Delta\eta,\Delta\phi)$ was obtained for charged hadrons produced in strong and electromagnetic processes in Be+Be interactions within the \NASixtyOne acceptance. The acceptance maps are available in Ref.~\cite{ppm_edms}.
        
\section{Experimental setup}\label{sec:setup}

This section gives a brief description of the experimental setup used for recording Be+Be collisions. 

\subsection{Detector}

\begin{figure*}
  \centering
  \includegraphics[width=0.8\textwidth]{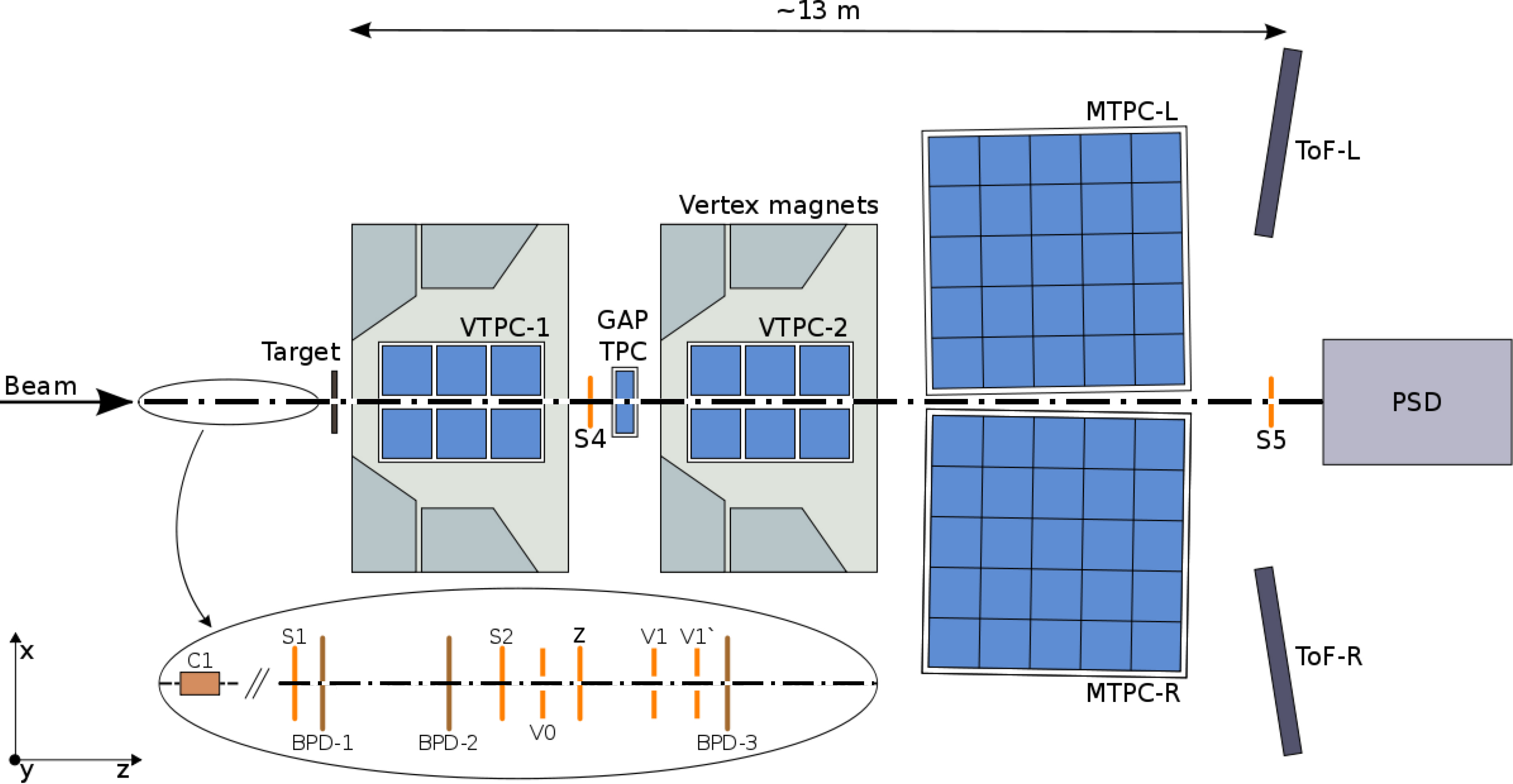}
  \caption{
    (Color online) The schematic layout of the \NASixtyOne experiment at the CERN SPS ~\cite{Abgrall:2014fa} showing the components used for the Be+Be energy scan (horizontal cut, not to scale). The beam instrumentation is sketched in the inset. Alignment of the chosen coordinate system as shown in the figure; its origin lies in the middle of VTPC-2, on the beam axis. The nominal beam direction is along the $z$ axis. The magnetic field bends charged particle trajectories in the $x$--$z$ (horizontal) plane. The drift direction in the TPCs is along the $y$ (vertical) axis.}
  \label{fig:detector-setup}
\end{figure*}

The \NASixtyOne experiment is a multi-purpose facility designed to measure particle production in
nucleus-nucleus, hadron-nucleus and proton-proton interactions~\cite{Abgrall:2014fa}. The detector
is situated at the CERN Super Proton Synchrotron (SPS) in the H2 beamline of the North experimental area.
A schematic diagram of the setup is shown in Fig.~\ref{fig:detector-setup}.
The main components of the produced particle detection system are four large volume
Time Projection Chambers (TPC). Two of them, called Vertex TPCs (VTPC), are located
downstream of the target inside superconducting magnets with maximum combined bending power of 9~Tm.
The magnetic field was scaled down in proportion to the beam momentum in order to obtain
similar phase space acceptance at all energies. The main TPCs (MTPC) and two walls of
pixel Time-of-Flight (ToF-L/R) detectors are placed symmetrically to the beamline downstream of the magnets.
The fifth small TPC (GAP TPC) is located between VTPC-1 and VTPC-2 directly on the beam line.
The TPCs are filled with Ar:CO$_{2}$ gas mixtures in proportions 90:10 for the VTPCs and the GAP-TPC, 
and 95:5 for the MTPCs. 

The Projectile Spectator Detector (PSD), which measures mainly the energy of projectile spectators,
is positioned 20.5 m (16.7 m) downstream of the target and behind the MTPCs at 75$A$ and 150\AGeVc
(19$A$, 30$A$, 40\AGeVc), centered in the transverse plane on
the deflected position of the beam. The PSD allows to select the centrality (violence) of the
collision by imposing an upper limit on the measured spectator energy. For more details see Secs.~\ref{sec:PSD} and~\ref{sec:centrality_determination}.

\begin{figure}[h]
\centering
\includegraphics[width=\textwidth]{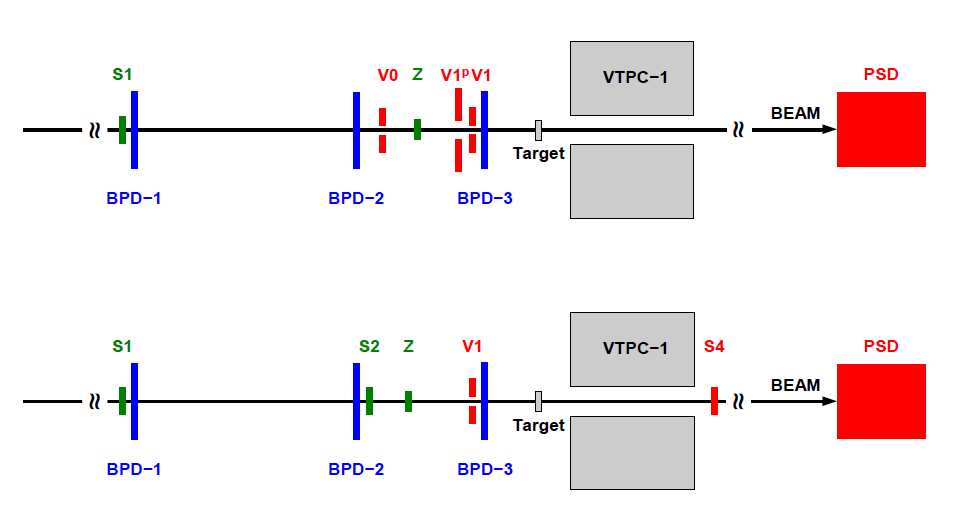}
\caption{(Color online) The schematic of the placement of the beam and trigger detectors in high momentum \emph{(top)} and 
         low momentum \emph{(bottom)} data taking.}
\label{fig:beamAndTriggerDetectors}
\end{figure}

The beamline instrumentation is schematically depicted in Fig.~\ref{fig:beamAndTriggerDetectors}.
A set of scintillation counters as well as Beam Position Detectors (BPDs) upstream of the spectrometer
provide timing reference, selection, identification and precise measurement of the position
and direction of individual beam particles.

\subsection{$^9$Be Target}

The target was a plate of $^9$Be of 12~mm thickness placed 75 cm upstream of front face of VTPC-1. 
Mass concentrations of impurities were measured at 0.3\%
resulting in an estimated increase of the produced pion multiplicity by less than 0.5\%
due the small admixture of heavier elements~\cite{Banas:2018sak}. No correction was applied for
this negligible contamination. 
Data were taken with target inserted (90\% of all recorded events) and target removed (10\% of all recorded events). 

\subsection{$^7$Be Beam}

The beamline of \NASixtyOne is designed for obtaining high beam purity even with secondary ion beams.
The beam instrumentation (see Fig.~\ref{fig:beamAndTriggerDetectors}) consists of scintillator counters (S) 
used for triggering and beam particle identification, 
veto scintillation counters (V) with a hole in the middle for rejection of upstream interactions 
and beam halo particles, and a Cherenkov charge detector Z. Three 
Beam Position Detectors are used for determination of the charge of individual beam particles.

\begin{figure}[ht!]
        \centering
        \includegraphics[width=0.45\linewidth]{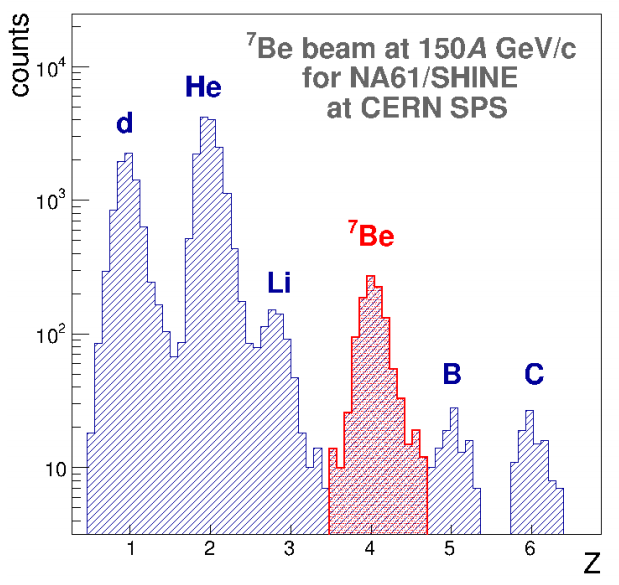}
        \caption{(Color online) Charge of the beam particles measured by the Z detector.}
        \label{fig:beamComposition}
\end{figure}

\begin{figure}[ht!]
        \centering
        \includegraphics[width=1\linewidth]{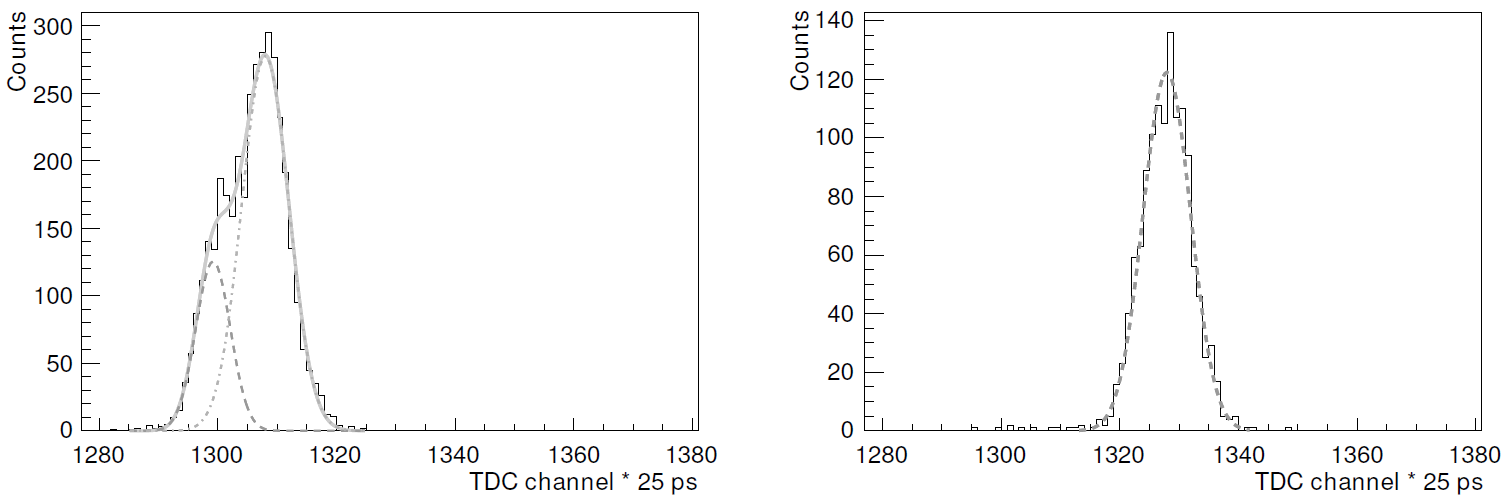}
        \caption{Time-of-flight of fragments of Z/A with momentum of 13.9~\GeVc. Left: carbon ions shows double Gaussian 
         structure due to two isotopes of carbon in the beam. Right: beryllium ions show single Gaussian distribution, 
         indicating isotopic purity of the beryllium in the beam.}
        \label{fig:beamAdet}
\end{figure}

This paragraph provides a brief description of the $^7$Be beam properties (see Ref.~\cite{Abgrall:7Bebeam}). 
Primary Pb$^{82+}$ ions extracted from the SPS were steered toward a 180 cm long beryllium fragmentation target 
placed 535~m upstream of the \NASixtyOne experiment. The result is a mixture of nuclear fragments consisting
of nucleons not participating in inelastic collisions (spectators) with momentum per nucleon 
equal to the beam momentum per nucleon smeared by the Fermi motion momentum. The spectrometers 
of the beamline allow to select beam particles based on the particle rigidity: $B\rho = 3.33 \cdot p_\text{beam}/Z$, 
where $B\rho$ can be adjusted by setting the current on the dipole magnets of the spectrometer 
and $p_\text{beam}$ is the momentum and $Z$ the charge of the beam particle.
Thus, the spectrometers select particles with the desired $A/Z$ ratio. The charge spectrum measured by
the Cherenkov Z detector for rigidity corresponding to $^7$Be is shown in Fig.~\ref{fig:beamComposition}. 
A well separated peak for charge $Z=4$ is visible and high purity $^7$Be ions can be selected
by a cut on the measured charge $Z$ as indicated by the red shading in Fig.~\ref{fig:beamComposition}. At test beam of momentum of 13.9\AGeVc it was also possible to
measure the time-of-flight of the beam particles. As demonstrated in Fig.~\ref{fig:beamAdet} the selected
fragments are high purity $^7$Be.

\subsection{Trigger}

The schematic of the placement of the beam and trigger detectors is shown in Fig.~\ref{fig:beamAndTriggerDetectors}. 
The trigger setup consists of a set of scintillation counters recording the presence of the beam particle (S1, S2), 
a set of veto scintillation counters with a hole used to reject beam particles passing far from the centre of 
the beamline (V0, V1), and a charge detector (Z). Beam particles were defined by the coincidence
T1 = $\text{S1} \cdot \text{S2} \cdot \overline{\text{V1}} \cdot \text{Z(Be)}$ and
T1 = $\text{S1} \cdot \overline{\text{V0}} \cdot \overline{\text{V1}} \cdot \overline{\text{V1'}} \cdot \text{Z(Be)}$ 
for low and high momentum data taking respectively. 
For the low beam momenta an interaction trigger detector (S4) was used to check whether the beam particle 
changed charge after passing through the target.
In addition, central collisions were selected by requiring an energy signal below a set threshold from the 16 central modules of the PSD (see Sec.~\ref{sec:PSD} for details). The event trigger condition thus was 
T2 = T1$\cdot \overline{\text{S4}} \cdot \overline{\text{PSD}}$ for 19$A$ and 30\AGeVc and 
T2 = T1$\cdot \overline{\text{PSD}}$ for 40$A$, 75$A$ and 150\AGeVc. 
The PSD threshold was set to retain from $\approx$~70\% to $\approx$~40\% of inelastic collisions 
at beam momenta of 19$A$ and 150\AGeVc, respectively. 

\subsection{The Projectile Spectator Detector}
\label{sec:PSD}

The centrality measurement for the events used in this report is based on information from 
the Projectile Spectator Detector (PSD), which is a modular compensating zero-degree calorimeter. 
Thanks to its modularity, there is only small dependence of the measured energy on the position 
of the particle and there is the possibility to determine centrality based on the energy 
measured by a subset of modules.

\begin{figure*}
\centering
\includegraphics[width=0.4\textwidth]{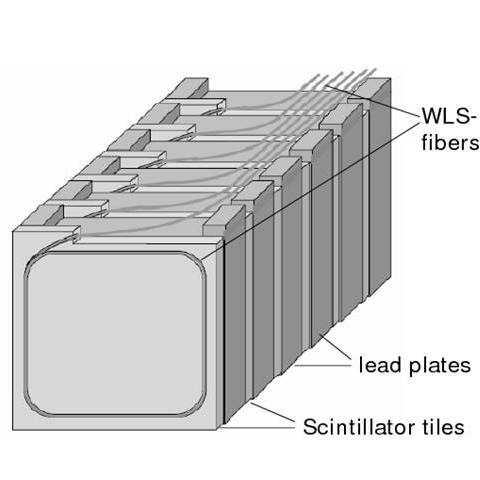}
\includegraphics[width=0.4\textwidth]{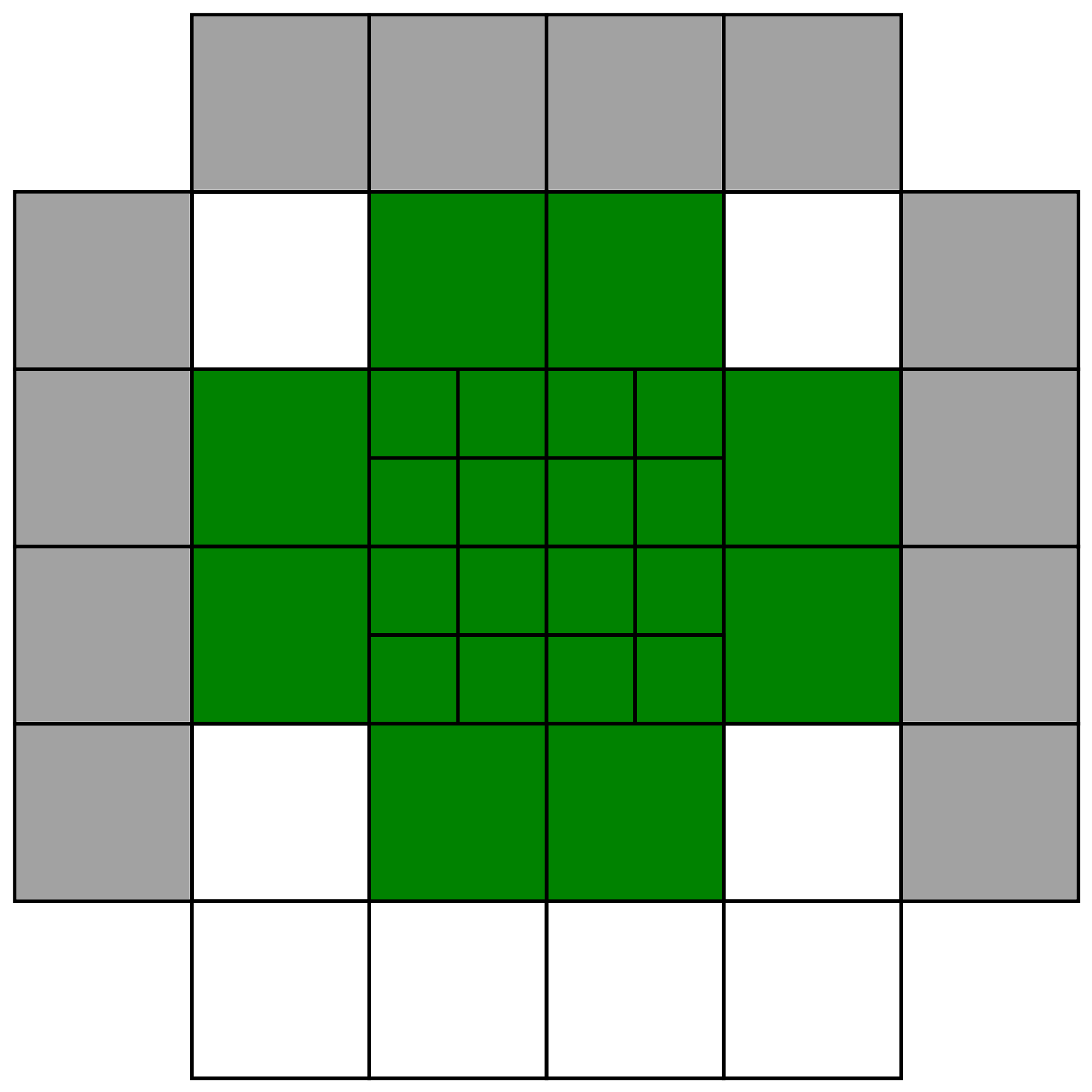}
\caption{(Color online) Left: construction schematic of a PSD module. Right: front face of the PSD showing modules 
in green used for centrality determination. The modules marked with gray colour were not installed
yet during taking at 40$A$, 75$A$ and 150\AGeVc.} 
\label{fig:PSD_selection}
\end{figure*}

The Projectile Spectator Detector used for this data taking consisted of 44 modules: 
16 small (10x10~cm) modules in the central region of the detector and 28 large (20x20~cm) 
modules placed around the small modules. Each PSD module consisted of 60 pairs of alternating 
plates of lead and scintillator (Fig.~\ref{fig:PSD_selection}, left). The signals from the scintillators 
of each module were read by 10 Silicon Photomultipliers (SiPMs). Each SiPM was connected through 
Wavelength Shifting (WLS) fibres to six consecutive scintillator plates in order to allow 
longitudinal calibration of the detector as well as the characterization of the longitudinal 
particle shower development. 

The event trigger placed a cut on the summed signals of the 16 small central modules. In the
offline analysis a subset of events was selected for which the summed energy from  
a larger group of modules (see Fig.~\ref{fig:PSD_selection}~right) was required to lie below 
a threshold value in order to select the 5\% most central collisions. The module choice was
based on the existence of an anti-correlation between the recorded energy and the 
charged particle multiplicity reconstructed in the TPCs (a description of this method 
as well as an acceptance map for the region of the PSD used for centrality selection 
can be found in Ref.~\cite{PSDAcceptanceMaps}). Moreover, the availability of
these modules was required at all beam momenta.

\subsection{Centrality determination}
\label{sec:centrality_determination}

The determination of centrality (violence) in \NASixtyOne is based on the measurement of the energy 
deposited in the PSD by spectators (the most central collisions correspond to a low number 
of spectators and therefore a low energy deposit in the PSD). Only the central 16 modules 
are included in the trigger. For offline analysis the energy recorded by a larger group of 
modules is summed (see Fig.~\ref{fig:PSD_selection}). 

\section{Data processing, simulation and detector performance}
\label{sec:data_proc}

Detector parameters were optimised by a data-based calibration procedure which also took into account their time dependence. Small adjustments were determined in consecutive steps for:
\begin{enumerate}[(i)]
  \item detector geometry, TPC drift velocities and distortions due to the magnetic field inhomogeneities in the corners of the VTPCs,
  \item magnetic field setting,
  \item specific energy loss measurements.
\end{enumerate}
Each step involved the reconstruction of data required to optimise a given set of calibration constants and time dependent corrections followed by verification procedures. Details of the procedure and quality assessment are presented in Ref.~\cite{Abgrall:1955138}. The resulting performance relative to the measurements of quantities relevant for this paper is discussed below.

The main steps of the data reconstruction procedure consists of: cluster finding in the TPC raw data, reconstruction of local track segments in each TPC separately, matching of local track segments and merging them into global tracks, track fitting based on a magnetic field map and determination of track parameters, determination of the interaction vertex using the beam information and the trajectories of tracks reconstructed in the TPCs, refitting of the particle trajectory using the interaction vertex as an additional point and determining the particle momentum vector at the interaction vertex.

A simulation of the \NASixtyOne detector response is used to correct the reconstructed data. For this purpose Be+Be collisions generated with the \Epos 1.99~\cite{Werner:2005jf,Kaptur:2015xzu} model were used to obtain the corrections for contamination by weak decays of strange particles, and reconstruction inefficiency of the \NASixtyOne detector.

The simulation consists of generating Be+Be collisions, propagating outgoing particles through the detector material using the GEANT 3.21 package~\cite{Geant3}, simulating the detector response using dedicated \NASixtyOne packages, simulating the interaction trigger selection, reconstructing the simulated events in the same way as the real data and matching reconstructed and simulated tracks based on the cluster positions (see Ref.~\cite{Abgrall:2011rma} for more details).

\section{Data selection and analysis}\label{sec:datasets}

This section describes the procedure used for the analysis. It consists of the following steps: application of event and particle selections, obtaining uncorrected experimental results and evaluation of correction factors based on simulations, and finally calculation of statistical uncertainties and estimation of systematic uncertainties.

\subsection{Event selection criteria}
\label{sec:event_cuts}

Due to the very small fraction of out-of-target interactions (less than one per mille) only interactions with target inserted were analysed, while target removed ones were not taken into account.
The events selected for the analysis reported in this paper had to satisfy the following conditions:

\begin{enumerate}[(i)]
\item event was selected by the central interaction trigger\footnote{Central interaction trigger accepted approximately 40-50\% minimum-bias events.} and was produced by a good quality\footnote{Good quality of the beam was assured by: ensuring that beam was composed purely from beryllium ions, rejection of off-time interactions in the target, proper positioning of beam along BPDs.} beam (Interactions).
\item event has a well-fitted main interaction vertex (Good vertex),
\item the maximal distance between the main vertex $z$ position and the centre of the beryllium target is 2.5~cm (Vertex $z$ pos.),
\item only the 5\% most central collisions (based on PSD spectator energy measurement) are accepted (Centrality 5\%).
\end{enumerate}

Table~\ref{tab:data_event_cuts} presents the number of events analysed for Be+Be reactions at five beam momenta.

\begin{table}
  \centering
\begin{tabular}{|l|c|c|c|c|}
\hline
	 & Interactions & Good vertex & Vertex $z$ pos. & Centrality 5\% \\
\hline
19\AGeVc & 565136 & 497624 (88\%) & 425346 (75\%) & 29531 (5\%)\\
30\AGeVc & 647662 & 592741 (92\%) & 513857 (79\%) & 38550 (6\%)\\
40\AGeVc & 1833013 & 881618 (48\%) & 776899 (42\%) & 109512 (6\%)\\
75\AGeVc & 2030413 & 927225 (46\%) & 822710 (41\%) & 92741 (5\%)\\
150\AGeVc & 1644127 & 833934 (51\%) & 732824 (45\%) & 81525 (5\%)\\
\hline
\end{tabular}

  \caption{Number of events before and after cuts. See text for explanation of the columns.}
  \label{tab:data_event_cuts}
\end{table}

\subsection{Track selection criteria and acceptance}
\label{sec:track_cuts}

The tracks selected for the analysis had to satisfy the following conditions:

\begin{enumerate}[(i)]
\item the track fit of this charged particle converged (Good track),
\item the total number of reconstructed points on the track should be at least 30 and, at the same time, the sum of the number of reconstructed points in VTPC-1 and VTPC-2 should be at least 15 or the number of reconstructed points in the GAP TPC should be at least five (TPC nPoints),
\item the ratio of total number of reconstructed points (np) on the track to the potential number of points (nmp) should be between 0.5 and 1.2\footnote{Due to uncertainty of the momentum fitting and the fitted interaction point, the np/nmp ratio values may exceed 1. Hence, the upper limit for the ratio was established as 1.2.} (np/nmp).
\item the distance between the track extrapolated to the interaction plane and the interaction point (impact parameter) should be smaller or equal to 4~cm in the horizontal (bending -- $b_{x}$) plane and 2~cm in the vertical (drift -- $b_{y}$) plane\footnote{Track impact point resolution depends on track multiplicity in the event as well as the method of vertex determination. Typically, it is at the level of 2~cm in $x$ and 1 cm in $y$ plane.} ($b_{x}$ \& $b_{y}$),
\item tracks with \dedx and total momentum values characteristic for electrons are rejected\footnote{See Ref.~\cite{Abgrall:2013qoa} for the details of this cut.} (No e$^{-}$/e$^{+}$).
\end{enumerate}

Numbers of tracks after consecutive selection cuts are presented in Table~\ref{tab:data_track_cuts}.

\begin{table}[h]
\begin{tabular}{|l|c|c|c|c|c|}
\hline
	 & Good track & TPC nPoints & np/nmp & $b_{x}$ \& $b_{y}$ & No e$^{-}$/e$^{+}$ \\
\hline
19\AGeVc & 358504 & 257617 (72\%) & 234868 (66\%) & 227529 (63\%) & 215949 (60\%)\\
30\AGeVc & 605627 & 443356 (73\%) & 407537 (67\%) & 393572 (65\%) & 375621 (62\%)\\
40\AGeVc & 2009333 & 1499813 (75\%) & 1388029 (69\%) & 1338798 (67\%) & 1282954 (64\%)\\
75\AGeVc & 2413377 & 1823376 (76\%) & 1696029 (70\%) & 1640207 (68\%) & 1585880 (66\%)\\
150\AGeVc & 2943518 & 2185182 (74\%) & 2023686 (69\%) & 1962862 (67\%) & 1916593 (65\%)\\
\hline
\end{tabular}

 \caption{Number of tracks before and after cuts for the 5\% most central Be+Be collisions. See text for explanation of the columns.}
  \label{tab:data_track_cuts}
\end{table}

Model simulations were performed in $4\pi$ acceptance, thus the \NASixtyOne detector acceptance filter needed to be applied before comparisons with data. The detector acceptance was defined as a three-dimensional matrix ($p$,$p_T$,$\phi$) filled with 1 or 0 depending on whether the bin was or was not populated by particles reconstructed and accepted in the events (see Ref.~\cite{ppm_edms}).

For reconstructed tracks from simulation, the cut rejecting electrons and positrons was implemented differently. Due to the lack of information on simulated specific energy loss for reconstructed simulated tracks, a procedure called "matching" was introduced. It connects the currently examined reconstructed to the best matched simulated track candidate by comparing properties of simulated tracks before reconstruction with the properties of the simulated track after reconstruction. The selection of the best candidate is performed in two steps:
\begin{enumerate}
  \item Pre-select the candidates that have a minimum value for the matching ratio $A/B$, where $A$ is a number of points common for both simulated and reconstructed track and $B$ is number of points for the simulated track. The matching of points between simulated and reconstructed tracks is based on the respective positions of the points,
  \item From the preselected candidates, choose the one with the highest $A$ value.
\end{enumerate}

 During the process of the analysis it was found that the minimal matching ratio value has an impact on the magnitude of the correlation in the region near $(\Delta\eta,\Delta\phi) = (0,0)$. An analysis based on loss of accepted tracks versus increasing matching ratio value resulted in the choice of the optimal ratio of $A/B \geq 0.6$. The variation related to the choice of the minimal ratio was included in the systematic uncertainty (see Sec.~\ref{sec:systematic}).

\subsection{Corrections}
\label{sec:corrections}

In order to correct the results for biases due to off-line event and track selection, detection efficiency, contribution of weak decays and secondary interaction products, an identical procedure was applied to the simulated data. The \Epos 1.99 model was used for event generation as it was done for inelastic p+p interactions~\cite{Aduszkiewicz:2016mww}. Correction factors $\text{Corr}(\Delta\eta,\Delta\phi)$ were calculated bin-by-bin as the ratio of the correlation functions for simulated events (``pure'') and for the same events after processing through \Geant 3.21 \cite{Geant3} detector simulation and reconstruction (``rec''), filtered using the same event and track selection cuts as for the data.
The correlation function is derived from differences of extensive quantities of the two particles and 
is therefore not expected to be sensitive to the details of the centrality selection. Thus a special
correction is not required.
 
\subsection{Statistical uncertainties}

Statistical uncertainties of the correlation function are calculated in every $(\Delta\eta,\Delta\phi)$ bin using the following formula:
\begin{equation}
  \label{eq:sigma_C}
  \sigma(C) = \sqrt{\left [\text{Corr} \cdot \sigma(C^{\text{raw}}) \right]^2
    + \left[ C^{\text{raw}} \cdot \sigma(\text{Corr}) \right]^2},
\end{equation}
where $C^{\text{raw}}$ is the uncorrected correlation function obtained following Eq.~\ref{eq:correlations} and $\text{Corr}(\Delta\eta,\Delta\phi)$ is the correction factor (described in Sec.~\ref{sec:corrections}). Detailed evaluation of this formula is described in Sec.~4.3.1 in Ref.~\cite{BMthesis}.

In general, statistical uncertainties do not exceed 5\%. The highest uncertainties are for $\Delta\eta$ regions with lower statistics, i.e. for $\Delta\eta > 2$ and for positively and negatively charged pairs of particles and lower beam momenta. Below $\Delta\eta = 2$ statistical uncertainties are within 3\% for beam momenta 19$A$, and 30\AGeV and within 1.5\% for 40$A$, 75$A$ and 150\AGeV.

\subsection{Estimation of systematic uncertainties}
\label{sec:systematic}

In order to estimate systematic uncertainties, the data were analysed with loose and tight event and track selection cuts. By modifying cuts, one changes the magnitude of the corrections due to various biasing effects. If the simulation perfectly reproduces the data, corrected results should be independent of the cuts. A dependence on the selection criteria is due to imperfections of the simulation and is used as an estimate of the systematic uncertainty. For example, systematic uncertainty caused by weakly decaying particles is estimated by varying $b_x$ and $b_y$ cuts. The standard set of cut values, presented in Secs.~\ref{sec:event_cuts} and \ref{sec:track_cuts} together with values of loose and tight cuts are tabulated in Table~\ref{tab:standard_loose_tight}. 

\begin{table}
  \centering
  \begin{tabular}{|l|c|c|c|}
    \hline
    \multicolumn{4}{|c|}{\textbf{Event cuts}} \\
    \hline
    & \textbf{Loose} & \textbf{Standard} & \textbf{Tight} \\
    \hline
    \textbf{Interactions} & \multicolumn{3}{c|}{applied} \\
    \hline
    \textbf{Good vertex} & \multicolumn{3}{c|}{applied} \\
    \hline
    \textbf{Vertex $z$ pos.} & $\pm 10$~cm & $\pm 2.5$~cm & $\pm 1.25$~cm \\
    \hline
    \textbf{Centrality 5\%} & \multicolumn{3}{c|}{applied} \\
    \hline
    \multicolumn{4}{|c|}{\textbf{Track cuts}} \\
    \hline
    \textbf{Good track} & \multicolumn{3}{c|}{applied} \\
    \hline
    \textbf{Total TPC points} & $\geq 10$ & \multicolumn{2}{c|}{$\geq 30$} \\
    \hline
    \textbf{VTPC (GAP TPC) points} & $> 10(5)$ & $\geq 15(5)$ & $\geq 30(6)$ \\
    \hline
    \textbf{np/nmp} & $(0.5;1.2)$ & $(0.5;1.2)$ & $(0.7;1.0)$ \\
    \hline
    \textbf{$|b_x|$} & $\leq 6$~cm & $\leq 4$~cm & $\leq 0.8$~cm \\
    \hline
    \textbf{$|b_y|$} & $\leq 5$~cm & $\leq 2$~cm & $\leq 0.8$~cm \\
    \hline
    \textbf{$e^-/e^+$ cut} & \multicolumn{3}{c|}{applied} \\
    \hline
    \hline
    \textbf{Matching ratio (MC$^{\text{rec}}$ only)} & $\geq 0.5$ & $\geq 0.6$ & $\geq 0.7$ \\
    \hline
  \end{tabular}
  \caption{Event (top) and track (bottom) selection cuts. The standard cuts (centre) are used to obtain the final results, whereas the loose (left) and tight (right) cuts are employed to estimate systematic uncertainties (see Secs.~\ref{sec:event_cuts} and~\ref{sec:track_cuts}, respectively). The last row applies only to reconstructed tracks from simulation (MC$^{\text{rec}}$).}
  \label{tab:standard_loose_tight}
\end{table}

Results for both sets of cuts were subtracted bin-by-bin ($\text{loose} - \text{tight}$). Since the differences in all bins follow Gaussian distributions with mean close to 0, the systematic uncertainties were taken to be approximately equal to the standard deviation of the distribution. This procedure was performed for all charge combinations (all charge, unlike-sign, positively and negatively charge pairs) and for all beam momenta. Mean systematic uncertainties were calculated for two sub-regions of $\Delta\eta$. For $0 \leq \Delta\eta < 2$ mean systematic uncertainties are at the level 0.5\% for all charge combinations and beam momenta. For the region of $2 \leq \Delta\eta \leq 3$ the mean systematic uncertainties are higher and are up to 2\% for negatively charged pairs of particles at 19\AGeVc. Statistical uncertainties were not taken into account during this analysis.
\section{Results and discussion}\label{sec:results}

This section presents the final two-particle correlation results together with their possible explanations.

\subsection{Two-particle correlation function $C(\Delta\eta,\Delta\phi)$}
\label{sec:detadphi_results}

\begin{figure*}
  \centering
  \includegraphics[width=0.25\textwidth]{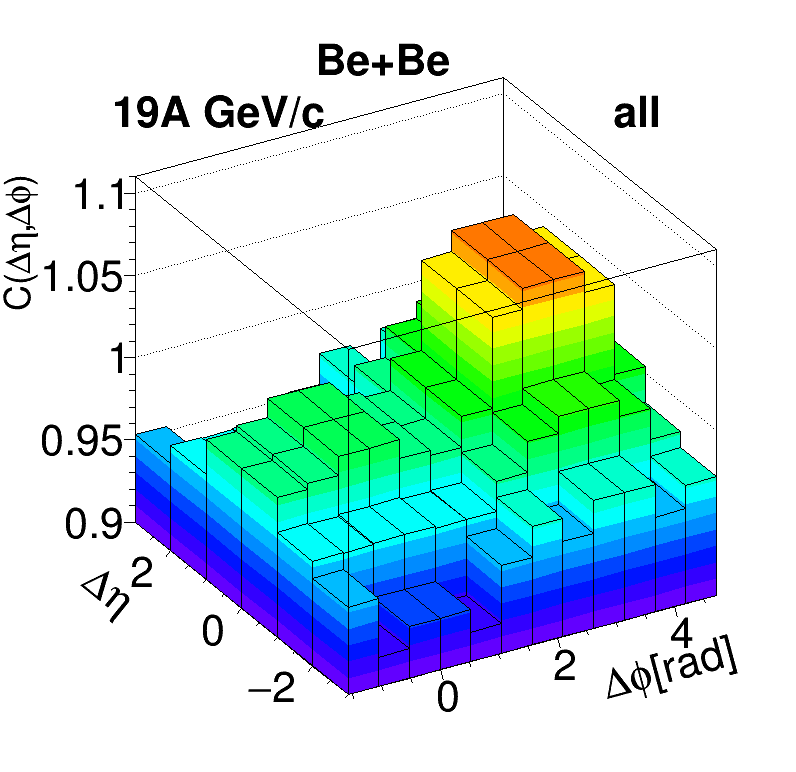}
  \includegraphics[width=0.25\textwidth]{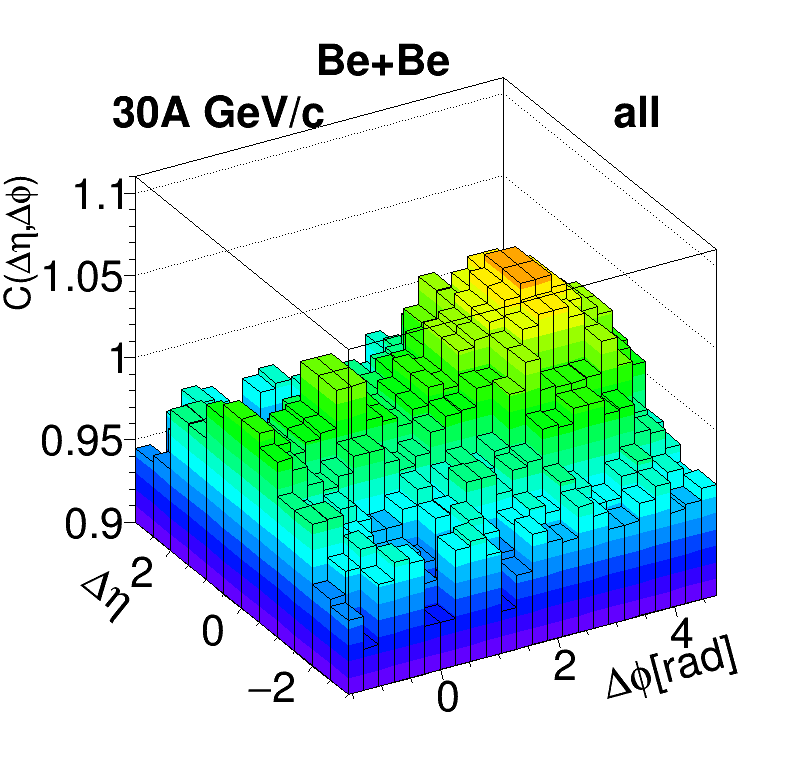}
  \includegraphics[width=0.25\textwidth]{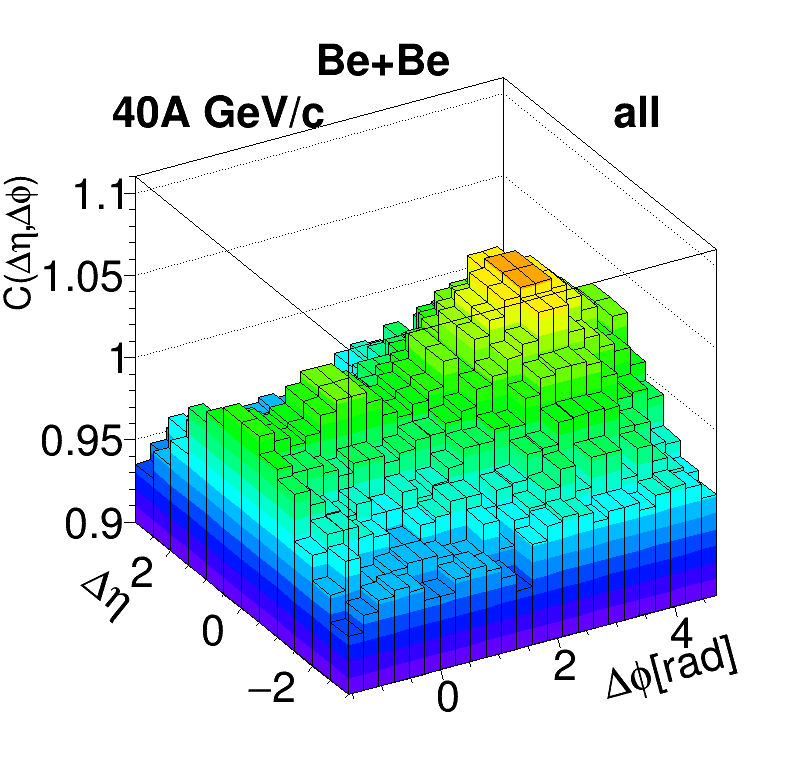}
  \\
  \includegraphics[width=0.25\textwidth]{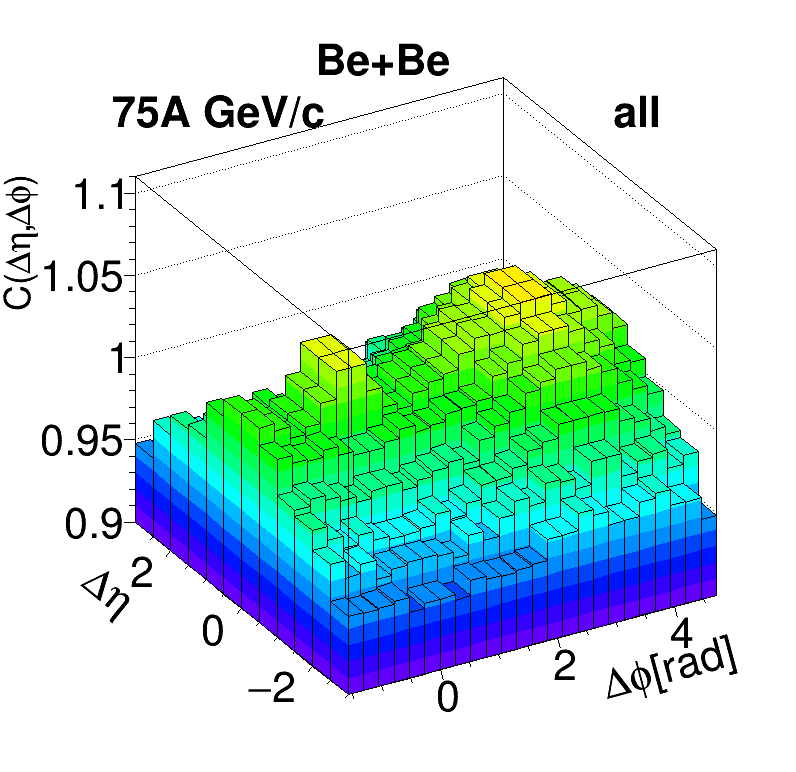}
  \includegraphics[width=0.25\textwidth]{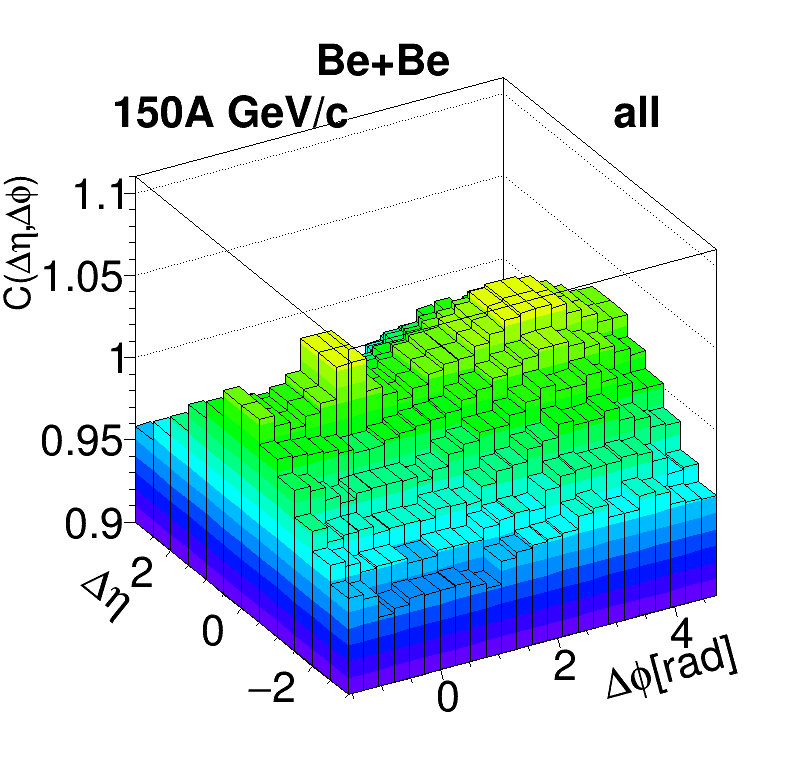}
  \caption{(Color online) Two-particle correlation function $C(\Delta\eta,\Delta\phi)$ for all charge pairs in the 5\% most central Be+Be collisions at 19$A$-150\AGeVc.}
  \label{fig:data_corr_all}
\end{figure*}

\begin{figure*}
  \centering
  \includegraphics[width=0.25\textwidth]{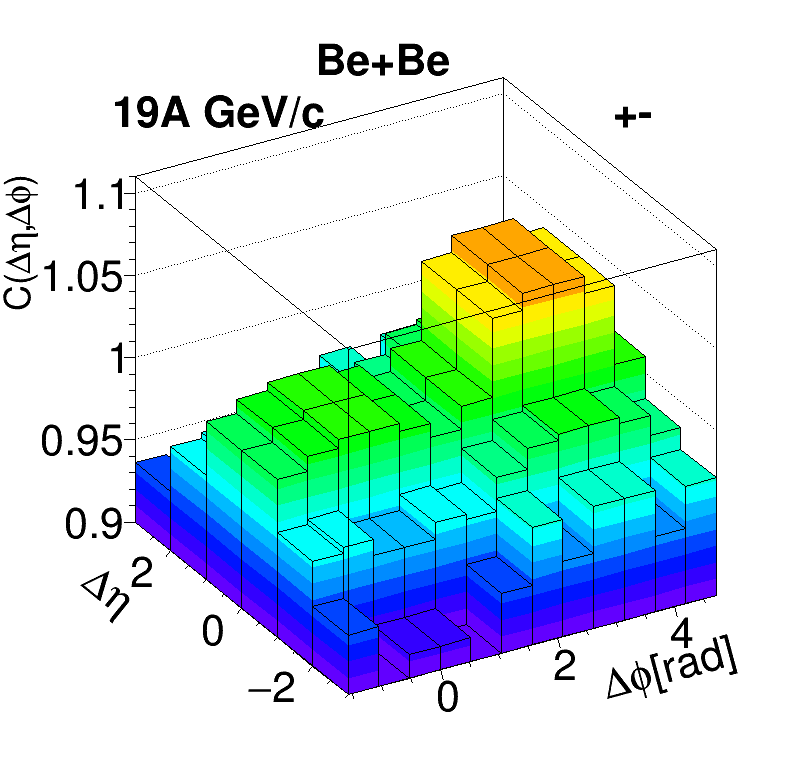}
  \includegraphics[width=0.25\textwidth]{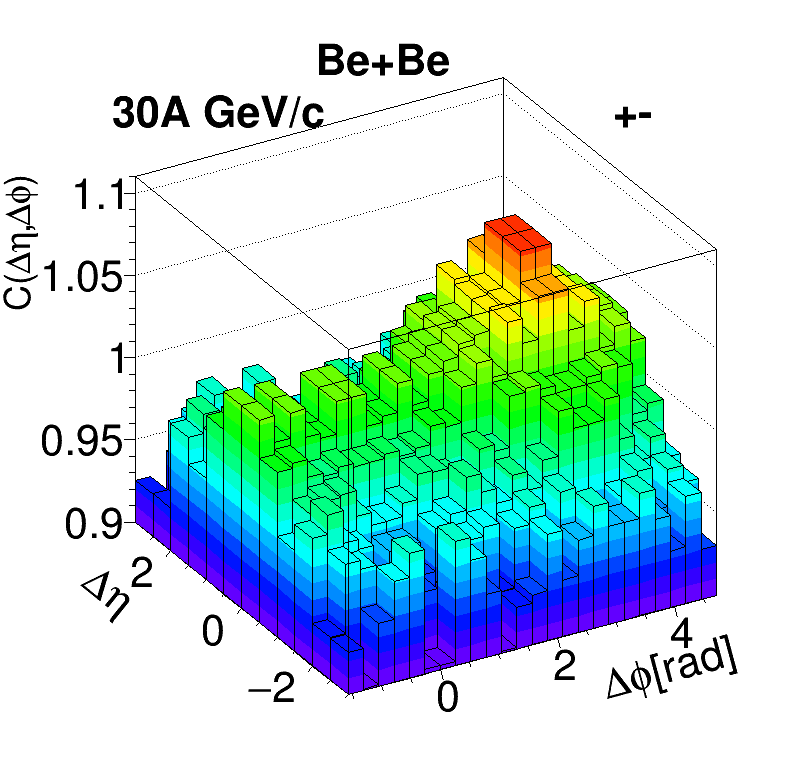}
  \includegraphics[width=0.25\textwidth]{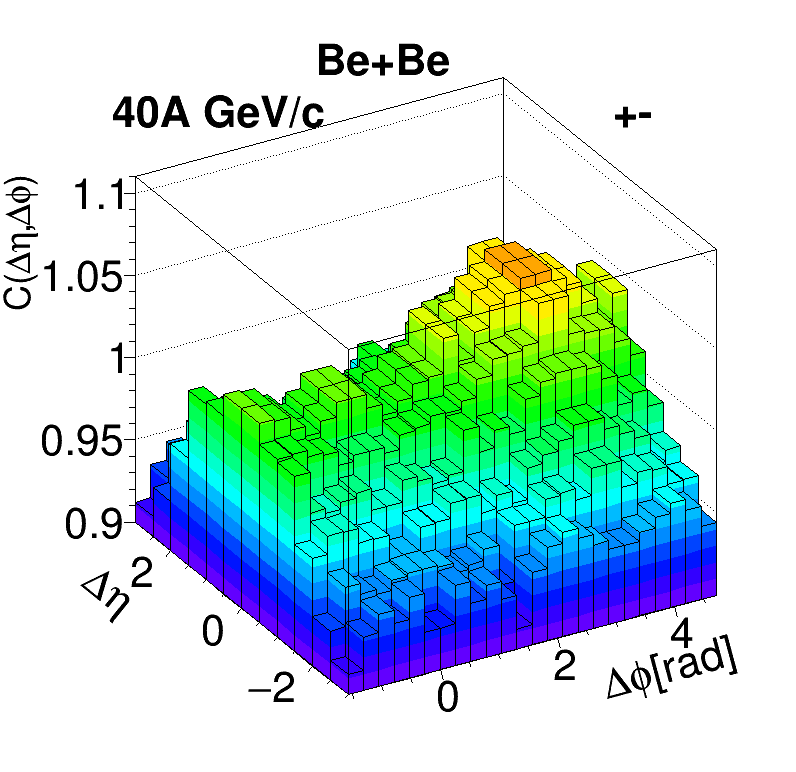}
  \\
  \includegraphics[width=0.25\textwidth]{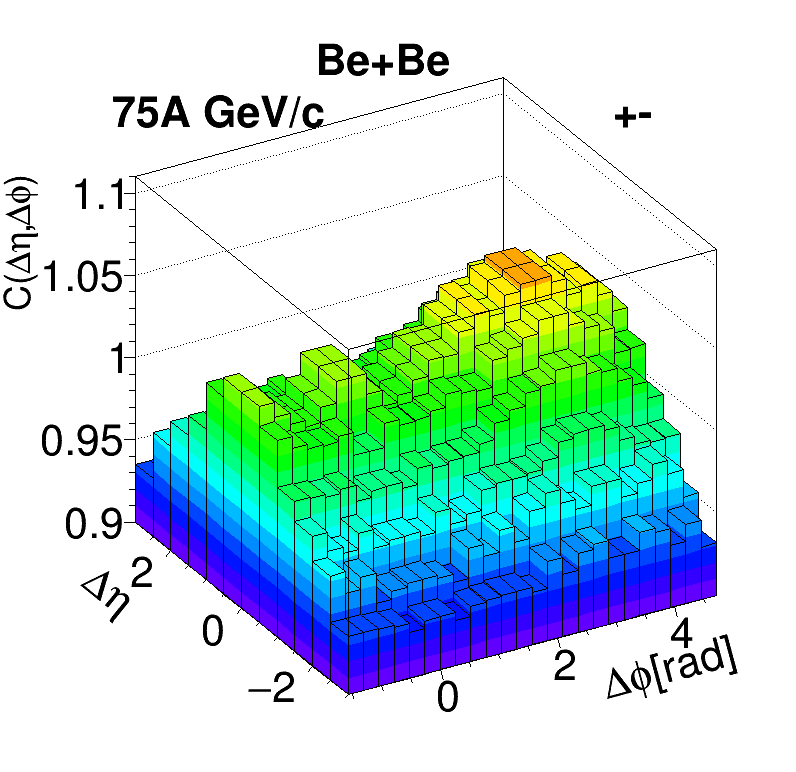}
  \includegraphics[width=0.25\textwidth]{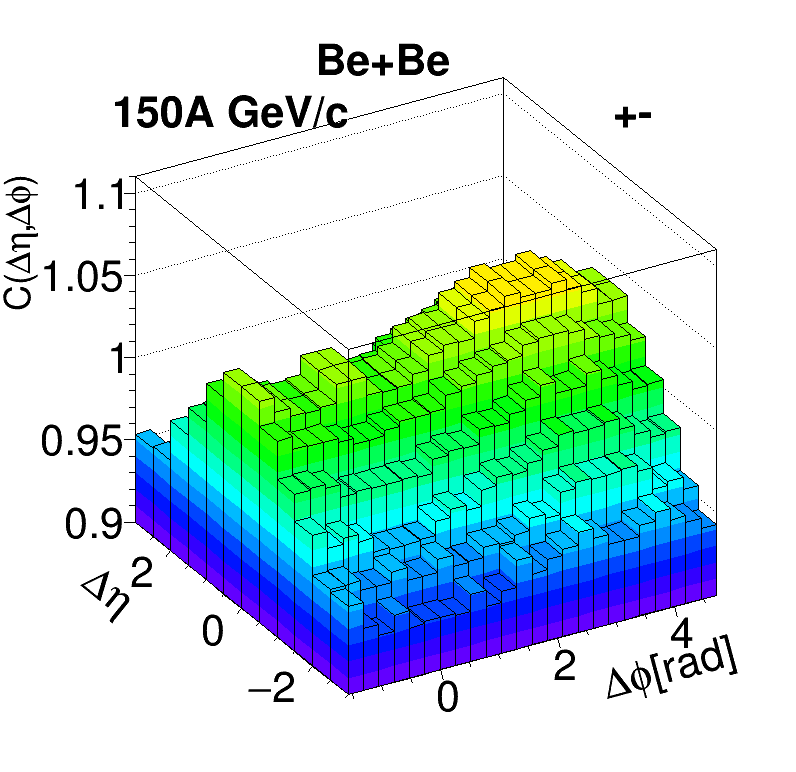}
  \caption{(Color online) Two-particle correlation function $C(\Delta\eta,\Delta\phi)$ for unlike-sign pairs in the 5\% most central Be+Be collisions at 19$A$-150\AGeVc.}
  \label{fig:data_corr_unlike}
\end{figure*}

\begin{figure*}
  \centering
  \includegraphics[width=0.25\textwidth]{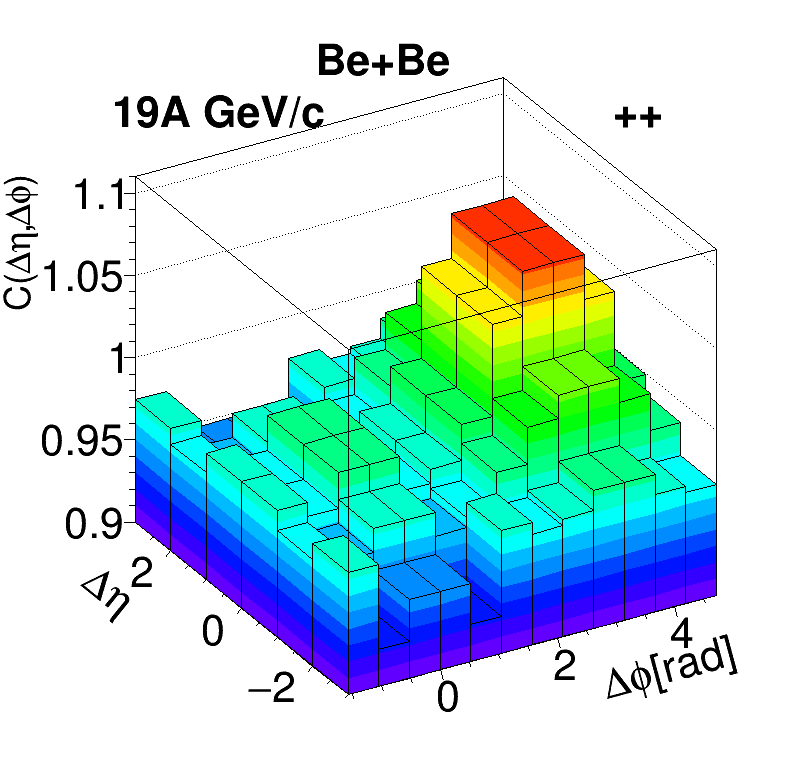}
  \includegraphics[width=0.25\textwidth]{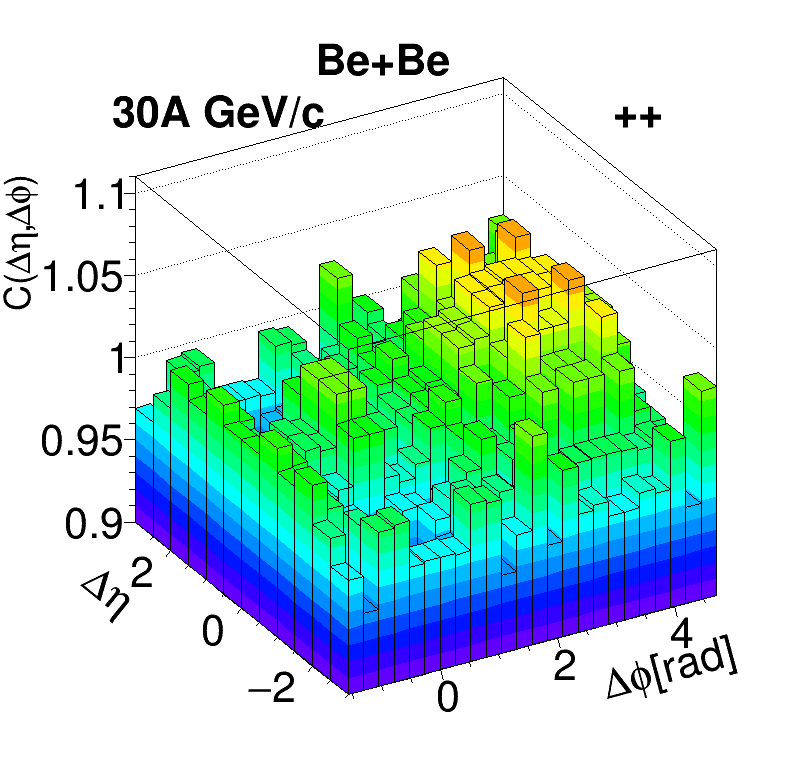}
  \includegraphics[width=0.25\textwidth]{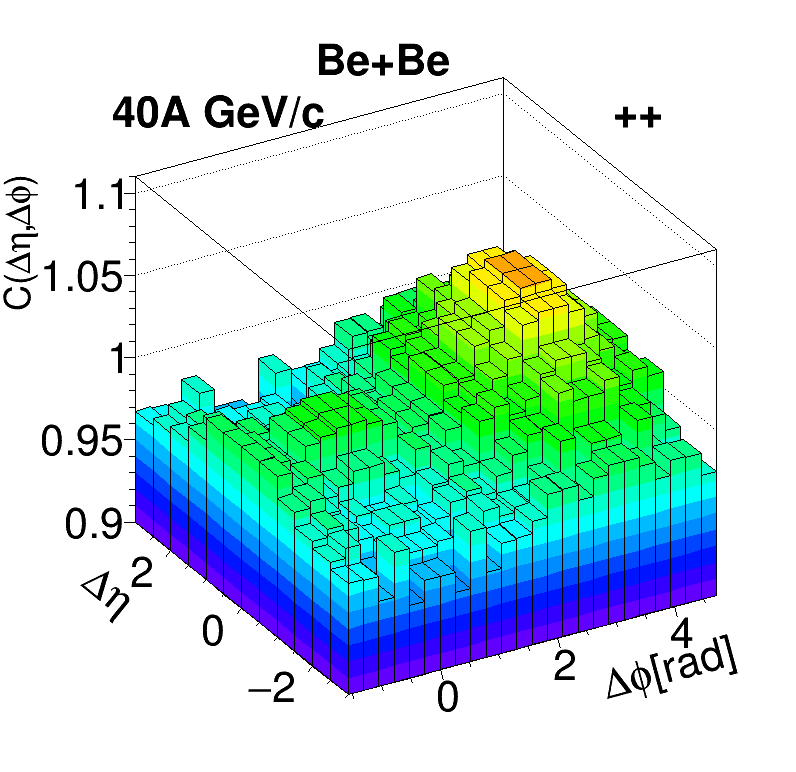}
  \\
  \includegraphics[width=0.25\textwidth]{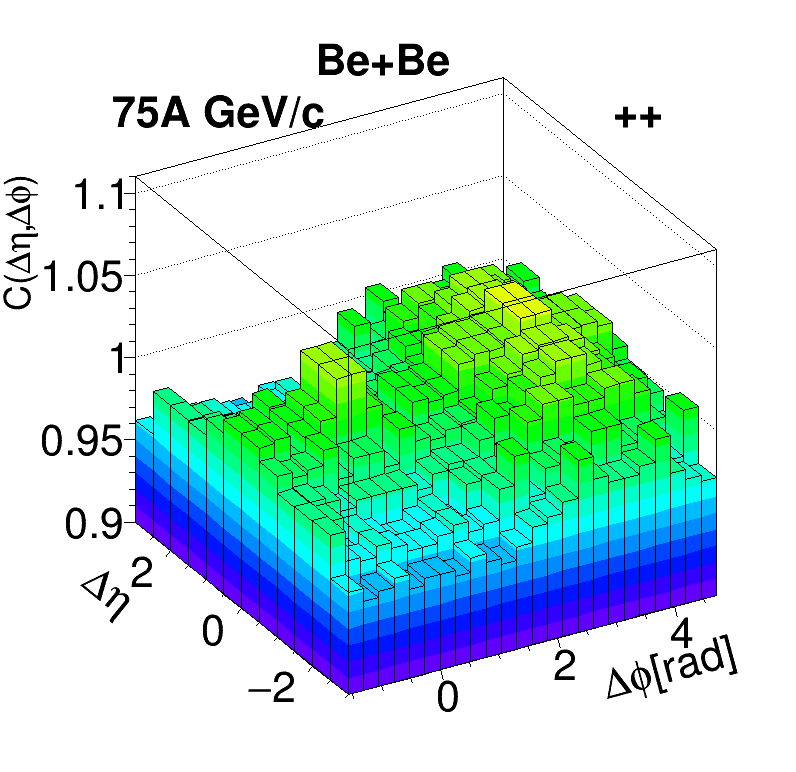}
  \includegraphics[width=0.25\textwidth]{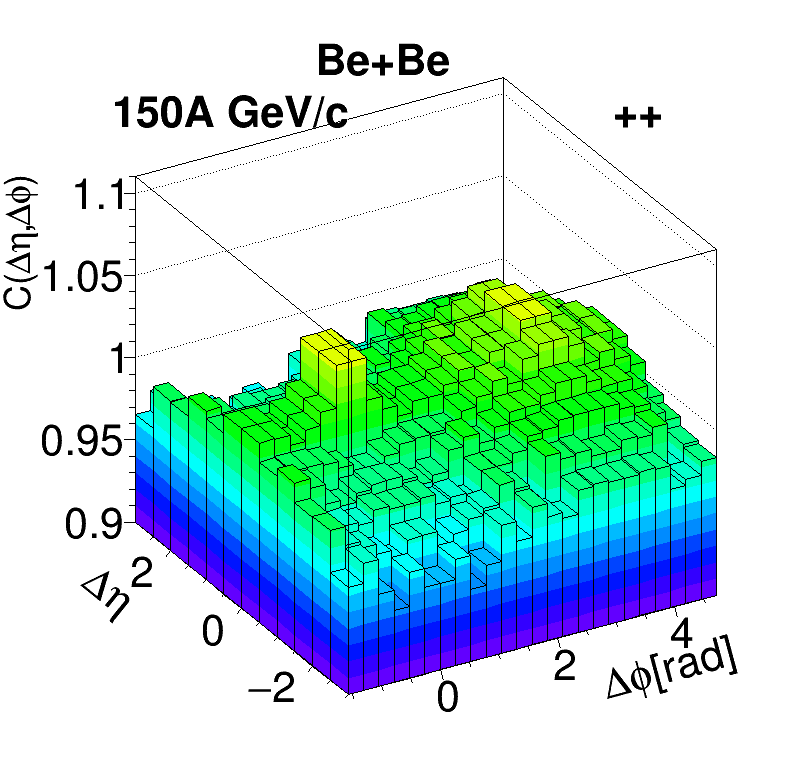}
  \caption{(Color online) Two-particle correlation function $C(\Delta\eta,\Delta\phi)$ for positive charge pairs in the 5\% most central Be+Be collisions at 19$A$-150\AGeVc.}
  \label{fig:data_corr_pos}
\end{figure*}

\begin{figure*}
  \centering
  \includegraphics[width=0.25\textwidth]{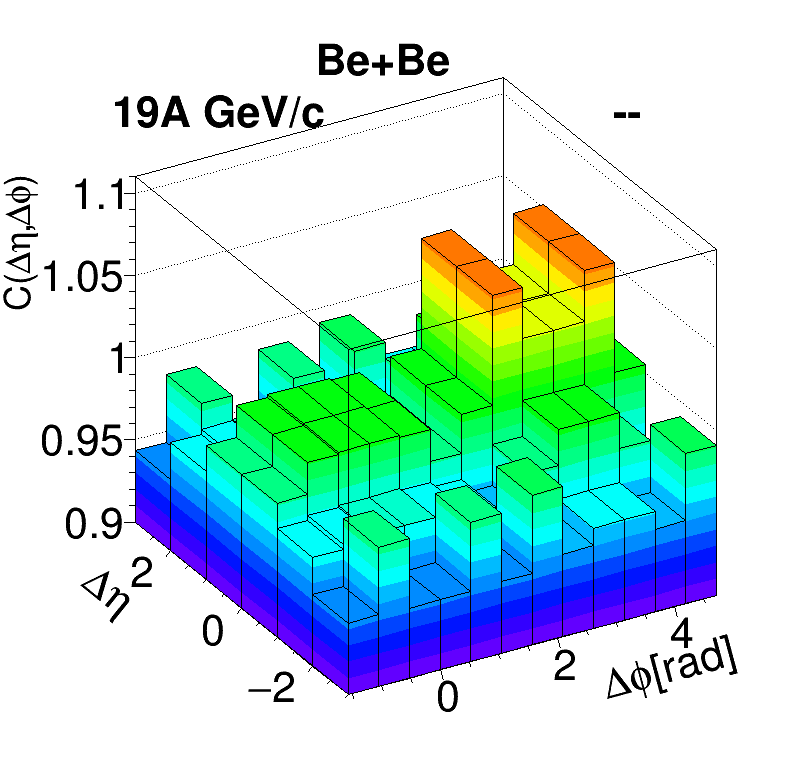}
  \includegraphics[width=0.25\textwidth]{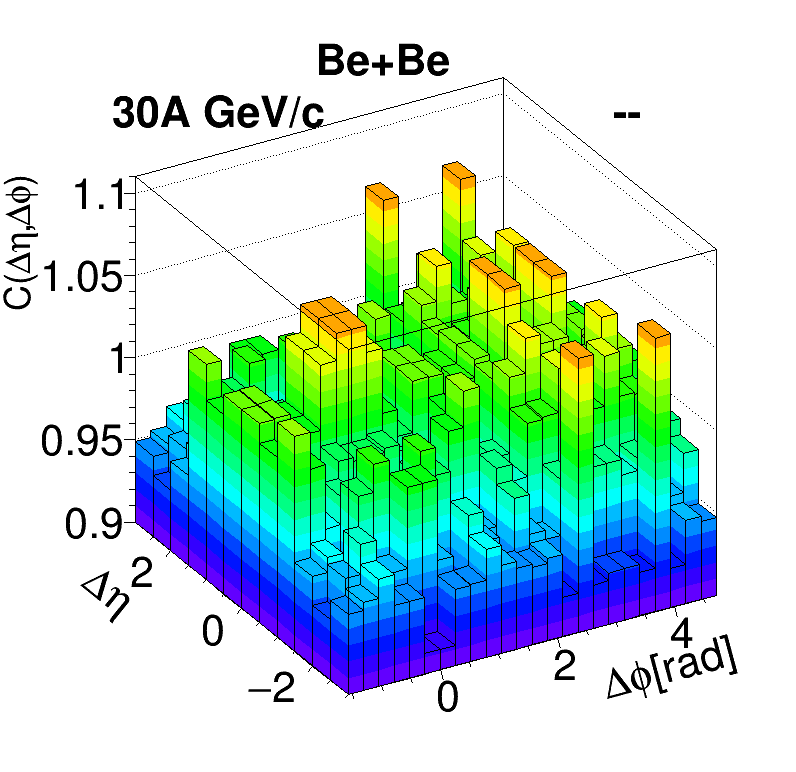}
  \includegraphics[width=0.25\textwidth]{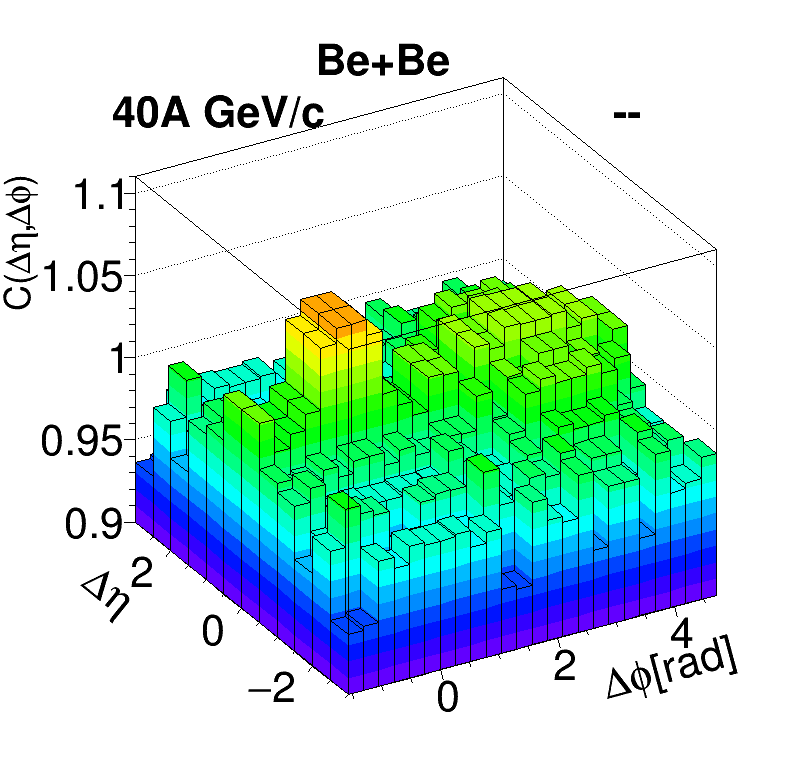}
  \\
  \includegraphics[width=0.25\textwidth]{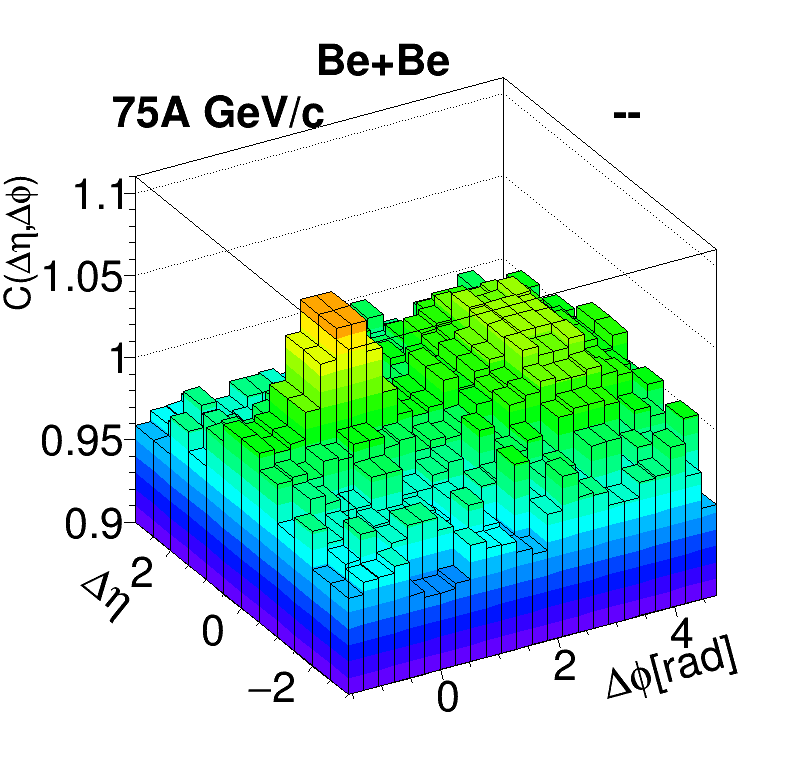}
  \includegraphics[width=0.25\textwidth]{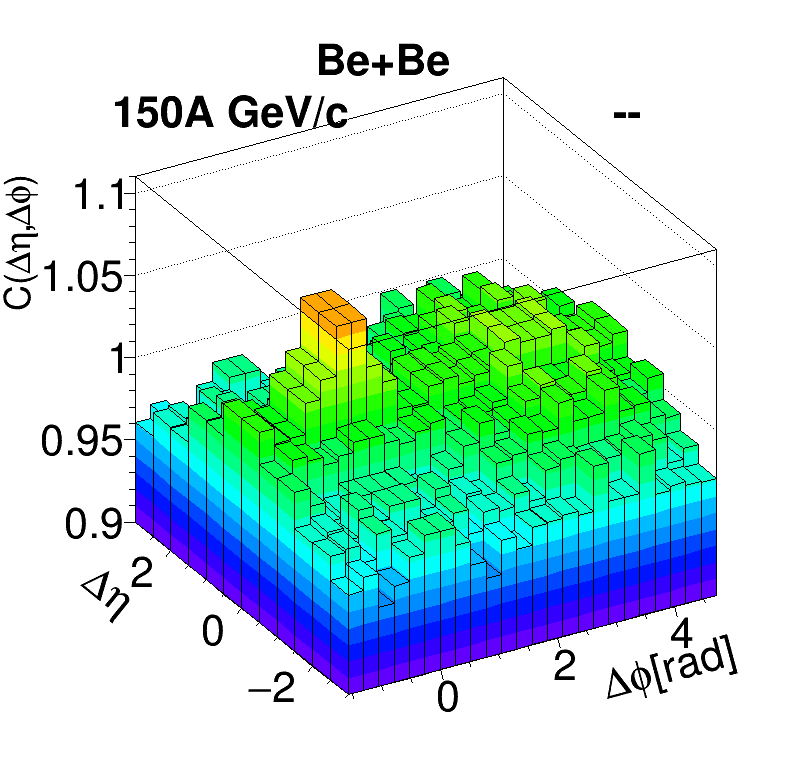}
  \caption{(Color online) Two-particle correlation function $C(\Delta\eta,\Delta\phi)$ for negative charge pairs in the 5\% most central Be+Be collisions at 19$A$-150\AGeVc.}
  \label{fig:data_corr_neg}
\end{figure*}

The corrected correlation functions for all charge pair combinations (all charge pairs, unlike-sign pairs, positively and negatively charge pairs) are presented in Figs.~\ref{fig:data_corr_all}, \ref{fig:data_corr_unlike}, \ref{fig:data_corr_pos} and \ref{fig:data_corr_neg}, respectively. Their values span the range between 0.9 and 1.1. Vanishing two-particle correlations would result in $C = 1$. 

Two-particle correlations in Be+Be collisions show the following features:
\begin{enumerate}[(i)]
  \item A maximum at $(\Delta\eta,\Delta\phi)=(0,\pi)$ emerging most probably due to resonance decays and momentum conservation. The maximum is most prominent for lower beam momenta and decreases with increasing beam momentum down to almost $C=1$ for 150\AGeVc for like-sign pairs. For unlike-sign pairs the maximum depends weakly on beam momentum. Comparing positive and negative charge pairs one can notice that in positive pairs the maximum is stronger. This may be explained by $\Delta^{++}$ resonance production and decay which contributes mostly to that correlation region. For negative charge pairs the maximum is barely visible due to the very low number of resonances decaying into two negatively charged particles.
  \item An enhancement  at $(\Delta\eta,\Delta\phi)=(0,0)$ likely due to a mix of different phenomena. It is rather broad (a $\Delta\phi$ bin corresponds here to $15^{\circ} \approx 0.26$~rad). At small $\Delta\phi$ (smaller than about $6^{\circ} \approx 0.1$~rad) it can be explained by a mixture of Quantum Statistic effects, Coulomb and final state interactions. A difference in height between positive and negative charge pairs is visible, namely in positive charge pairs the peak is significantly smaller than in negative charge pairs (especially for lower beam momenta). It is most probably due to the admixture of protons and an interplay between Bose-Einstein and Fermi-Dirac statistics. The HBT+Coulomb+FS correlations give a significant contribution to that region, however they probably do not explain all the excess. A more detailed discussion is provided in Sec.~\ref{sec:near_side_difference}.
\end{enumerate}

\section{Comparison with p+p data and with model predictions}\label{sec:models}

In this section two-particle correlation results presented in the previous section are compared with published \NASixtyOne results from p+p interactions~\cite{Aduszkiewicz:2016mww} and to theoretical predictions of the \Epos 1.99 and UrQMD 3.4~\cite{Bass:1998ca,Bleicher:1999xi} models.

\subsection{Comparison with correlations in p+p reactions}

\begin{figure*}
  \centering
  \includegraphics[width=0.25\textwidth]{figures/Data_corrected/150/corrected_all.png}
  \includegraphics[width=0.25\textwidth]{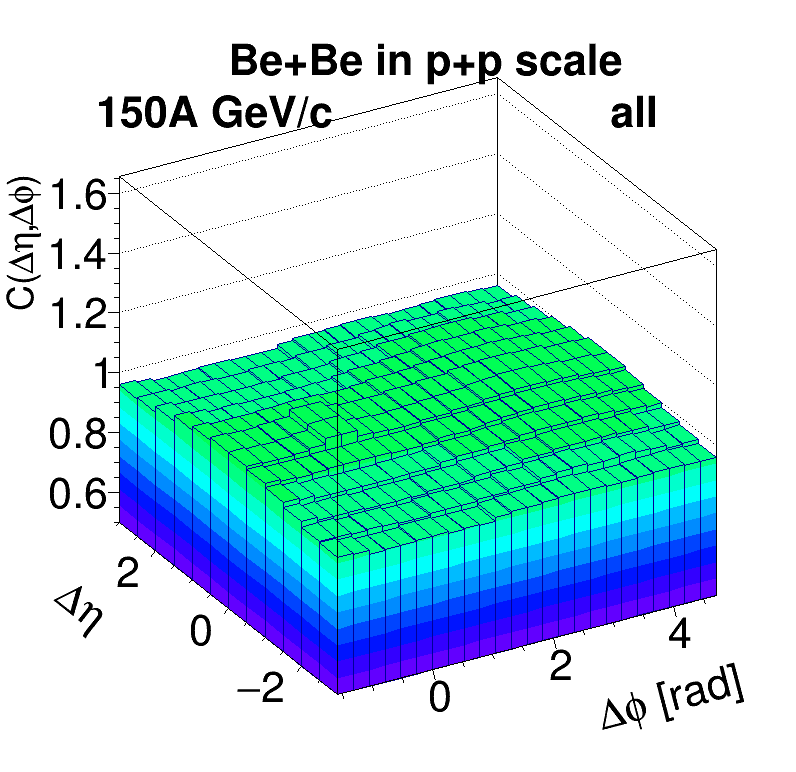}
  \includegraphics[width=0.25\textwidth]{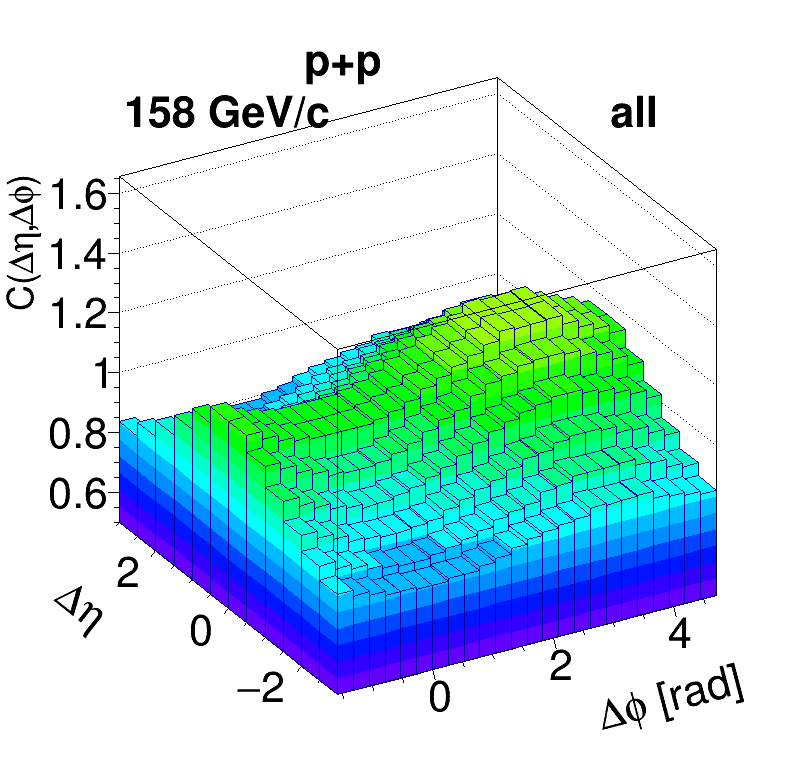}
  \caption{(Color online) Comparison of correlation functions $C(\Delta\phi)$ for the 5\% most central Be+Be collisions and inelastic p+p interactions. Left column shows the correlation function for all charge pairs in Be+Be collisions at beam momentum 150\AGeVc per nucleon. Middle column: the same Be+Be correlation function, but shown in the vertical scale used for p+p interactions. Right column: p+p correlation function for all pairs at beam momentum 158~\GeVc taken from Ref.~\cite{Aduszkiewicz:2016mww}.}
  \label{fig:BeBe_pp_optical_comparison}
\end{figure*}

This section presents a comparison of two-particle correlation measurements for the 5\% most central Be+Be collisions to those in inelastic p+p interactions reported in Ref.~\cite{Aduszkiewicz:2016mww}. Figure~\ref{fig:BeBe_pp_optical_comparison} shows an example comparison of results for beryllium-beryllium collisions (left panel) with results from proton-proton interactions (right panel). The most striking feature is the general difference in correlation strength. Due to the larger combinatorical background, the correlation strength is diluted in Be+Be collisions. For better comparison of the strengths the center panel of Fig.~\ref{fig:BeBe_pp_optical_comparison} presents the Be+Be results at the p+p scale. The dilution amounts to approximately a factor of 5 which is close to the ratio of pion multiplicities produced in Be+Be and p+p collisions. This is expected in simple models where e.g. the resonance to direct pion production ratio is assumed to be the same in both reactions.  

\subsubsection{The near-side correlations behaviour}
\label{sec:near_side_difference}
While both in Be+Be collisions and p+p interactions the away side hill is qualitatively similar, 
a visible difference of the structure at $C(\Delta\phi)$ can be seen. To visualize this, 
the ratio of the difference of the correlation functions from unity for Be+Be ($C^{\text{BeBe}}-1$) 
and p+p ($C^{\text{pp}}/5-1$) was calculated for all pair combinations and beam momenta 
using the following formula:

\begin{equation}
	\label{eq:ratio_BeBe_pp}
 	R^{\text{BeBe}}_{\text{pp}} = \frac{C^{\text{BeBe}}-1}{C^{\text{pp}}/5-1}.
\end{equation}

\begin{figure*}
  \centering
  \includegraphics[width=0.24\textwidth]{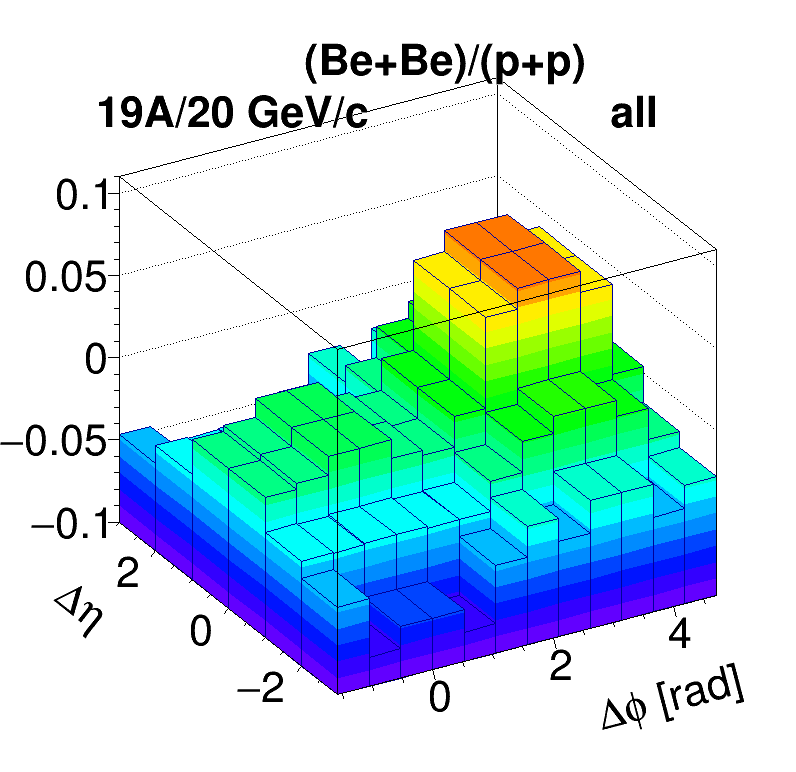}
  \includegraphics[width=0.24\textwidth]{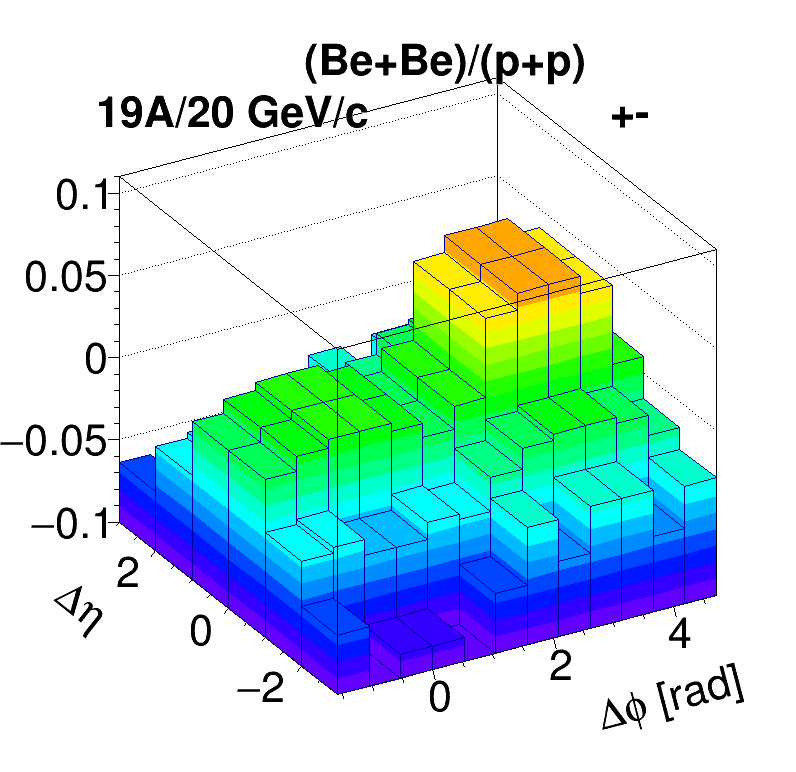}
  \includegraphics[width=0.24\textwidth]{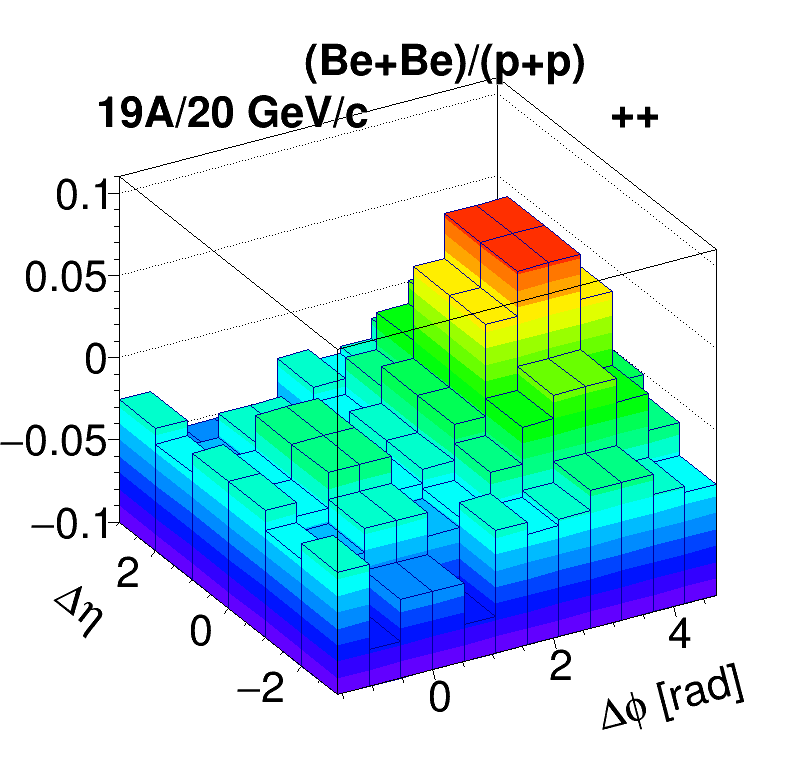}
  \includegraphics[width=0.24\textwidth]{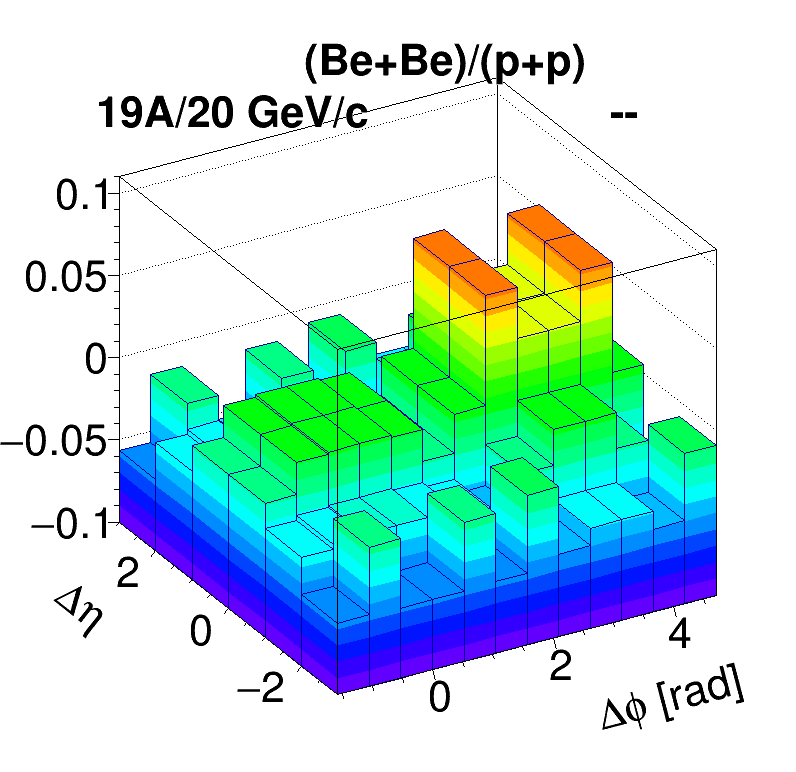}
  \\
  \includegraphics[width=0.24\textwidth]{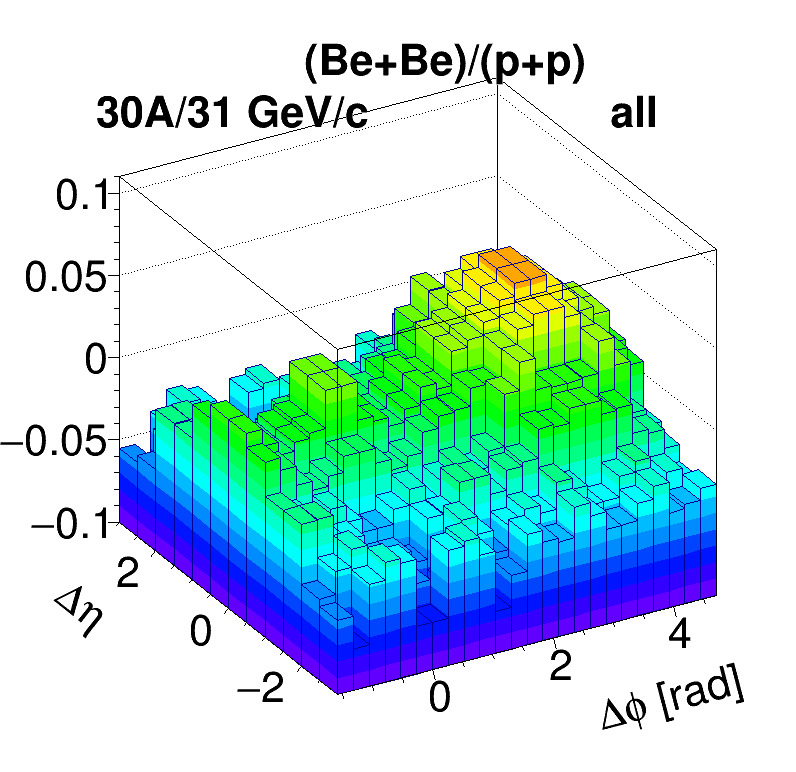}
  \includegraphics[width=0.24\textwidth]{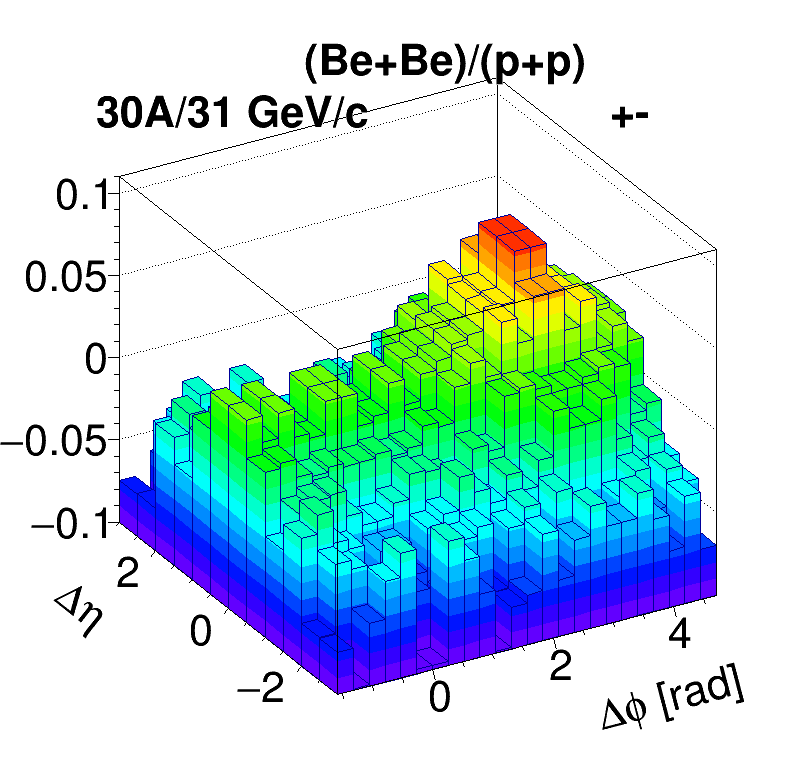}
  \includegraphics[width=0.24\textwidth]{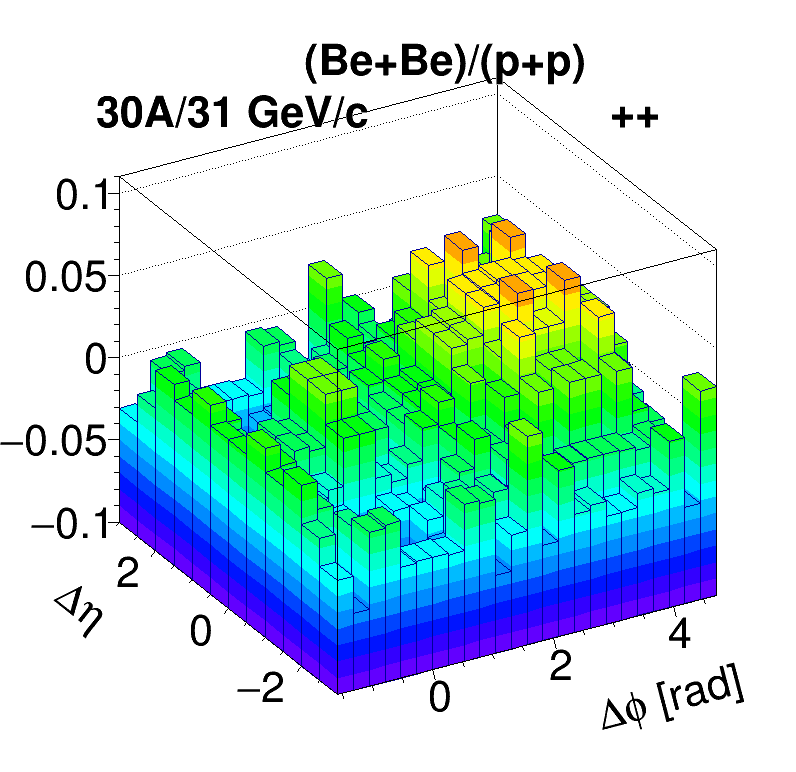}
  \includegraphics[width=0.24\textwidth]{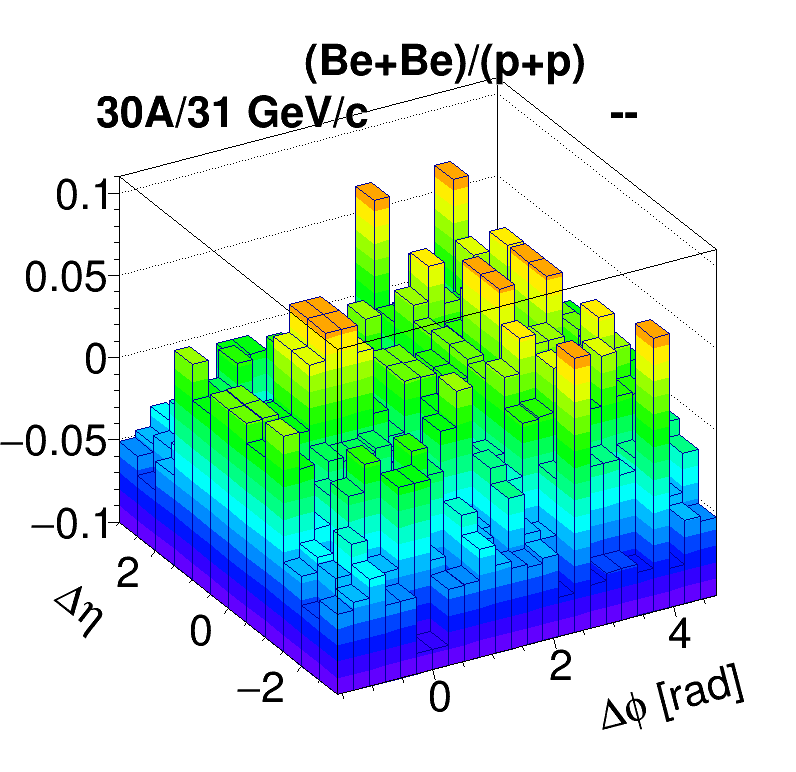}
  \\
  \includegraphics[width=0.24\textwidth]{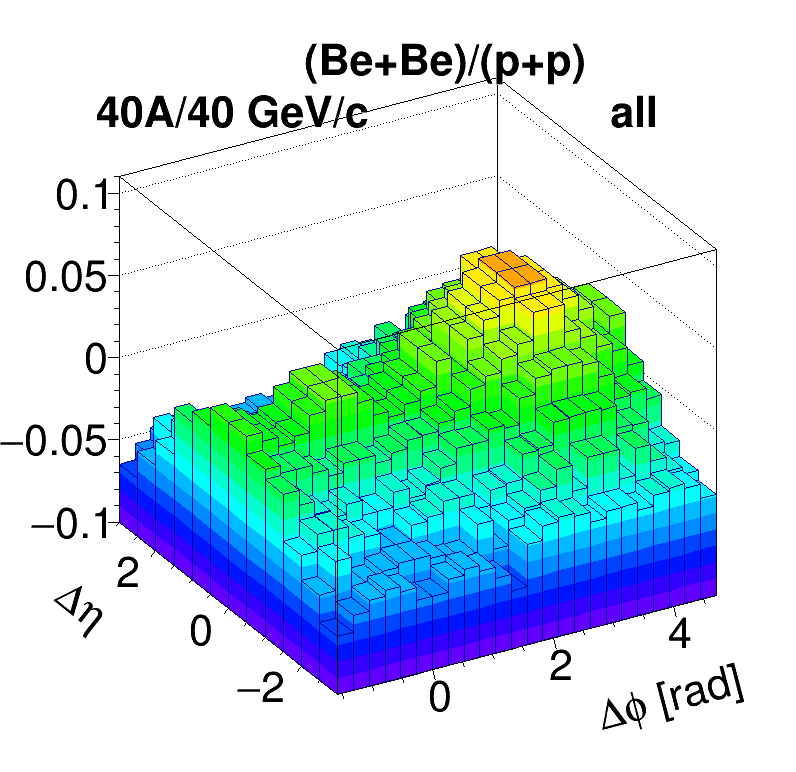}
  \includegraphics[width=0.24\textwidth]{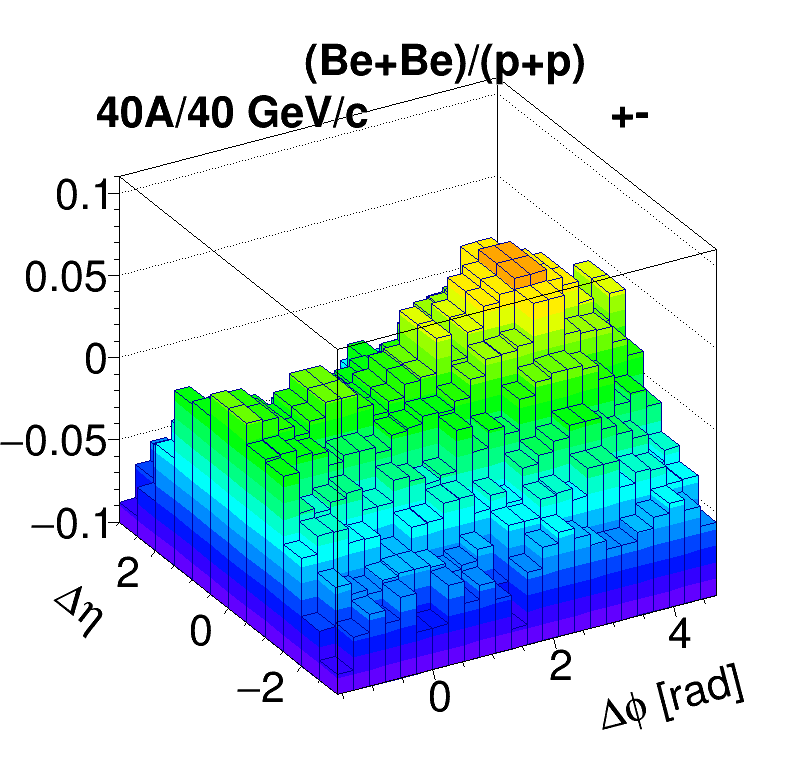}
  \includegraphics[width=0.24\textwidth]{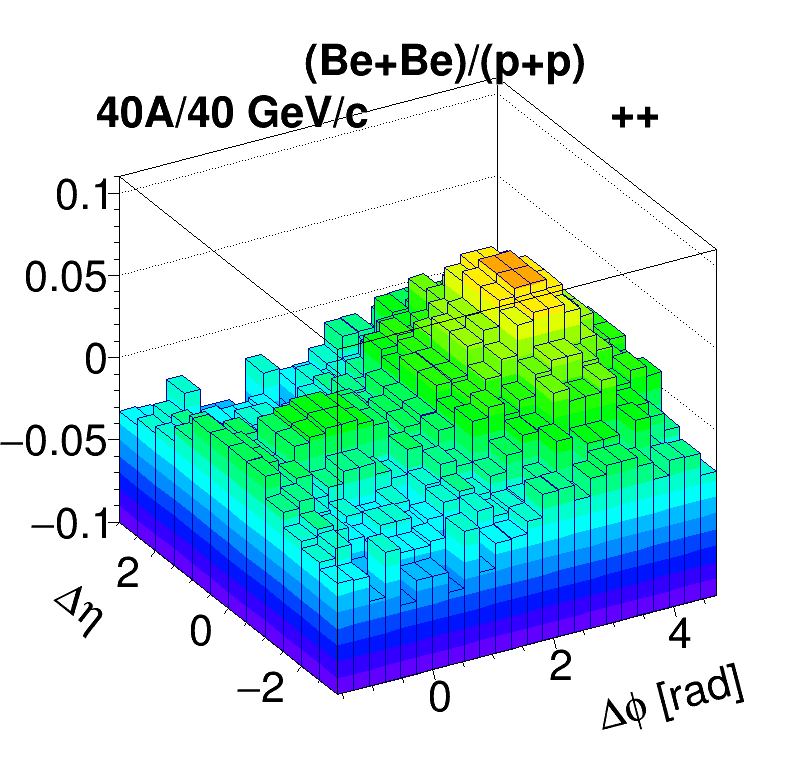}
  \includegraphics[width=0.24\textwidth]{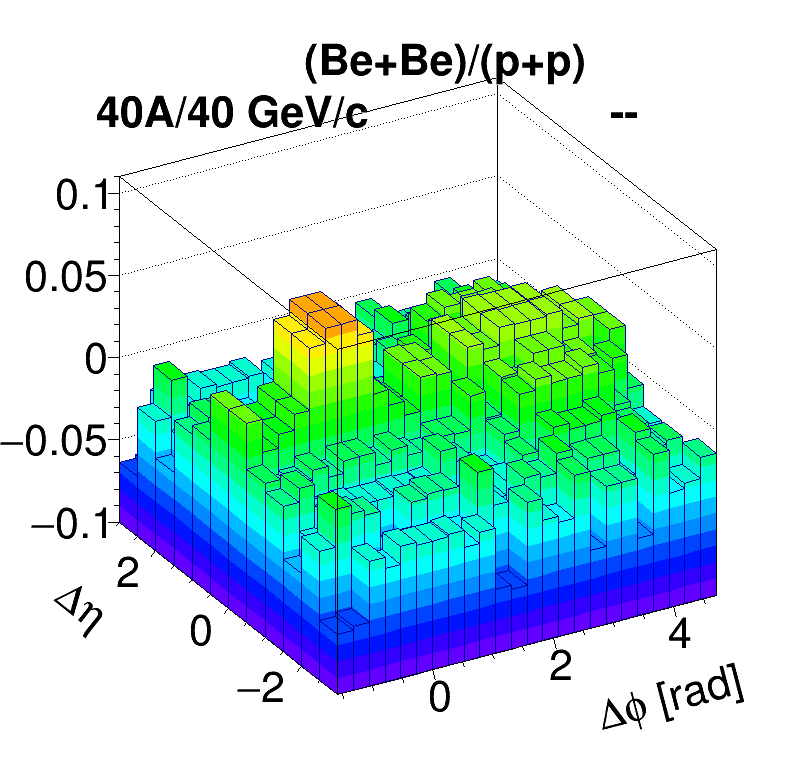}
  \\
  \includegraphics[width=0.24\textwidth]{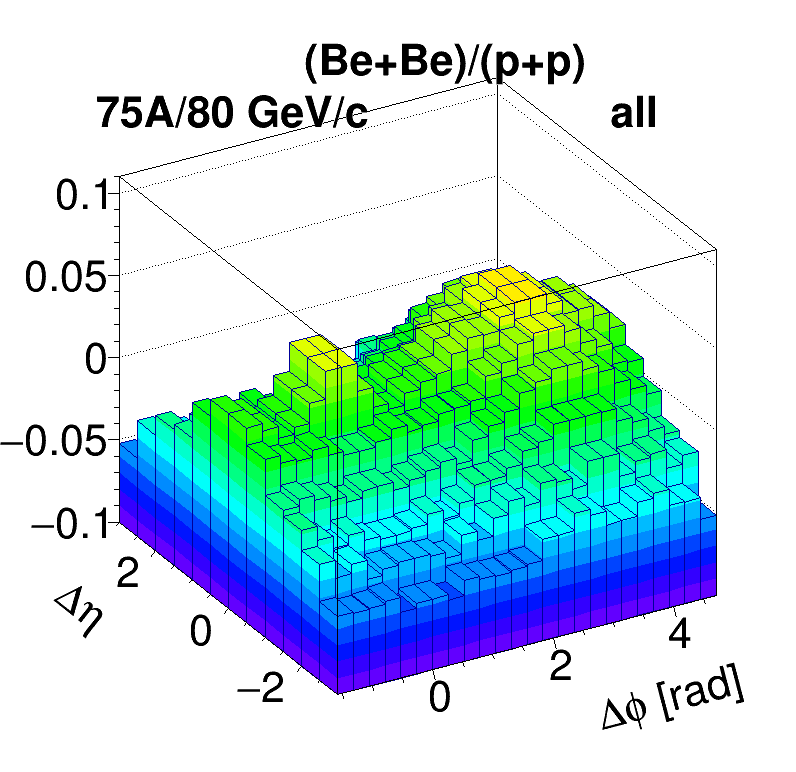}
  \includegraphics[width=0.24\textwidth]{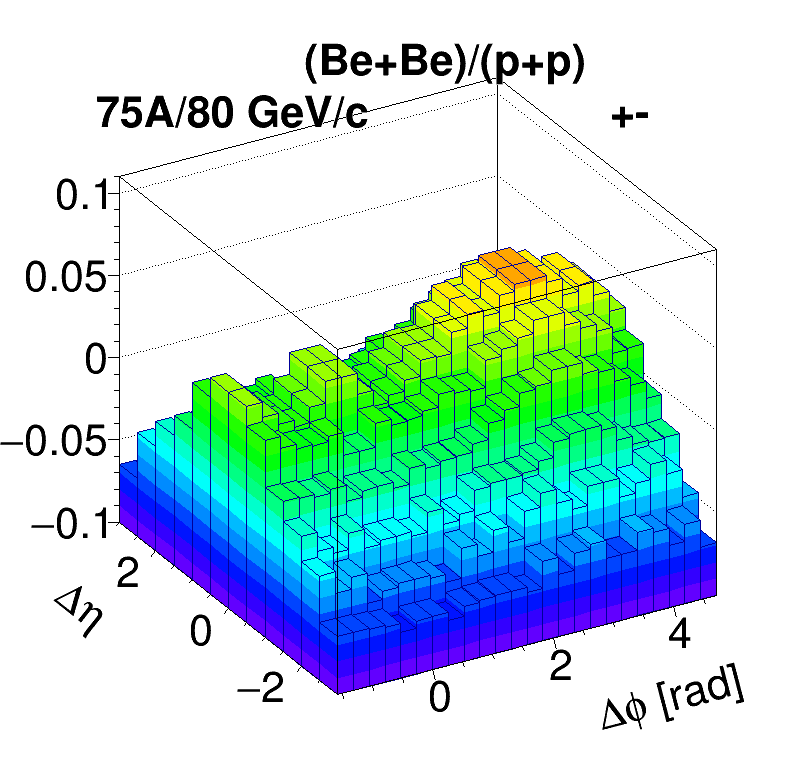}
  \includegraphics[width=0.24\textwidth]{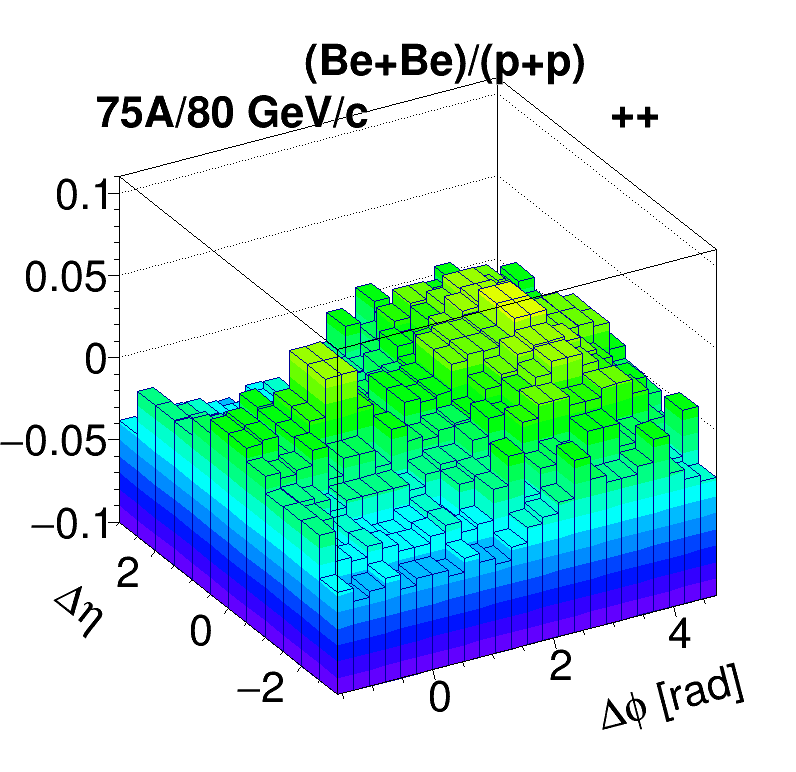}
  \includegraphics[width=0.24\textwidth]{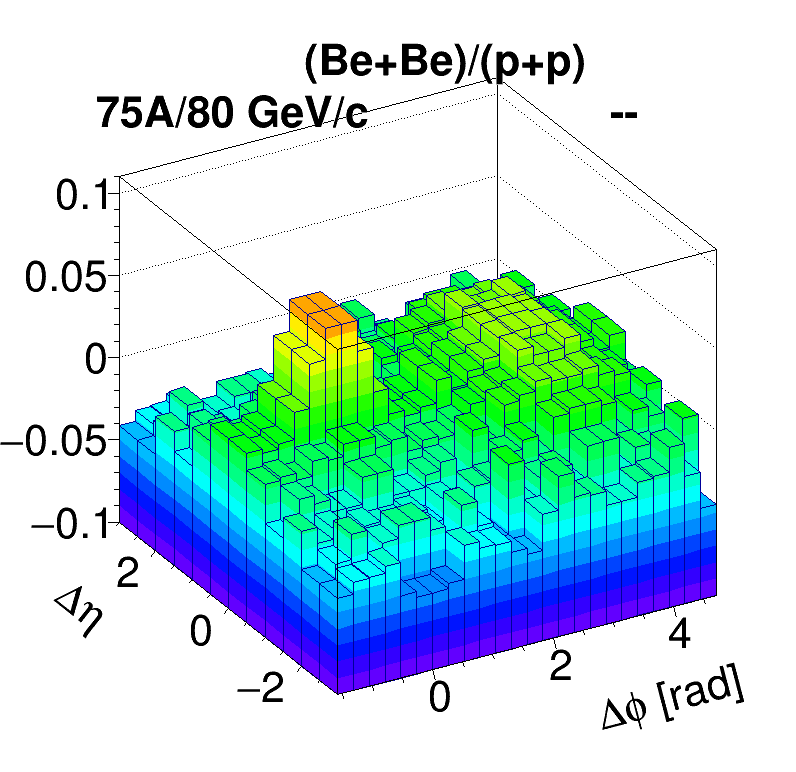}
  \\
  \includegraphics[width=0.24\textwidth]{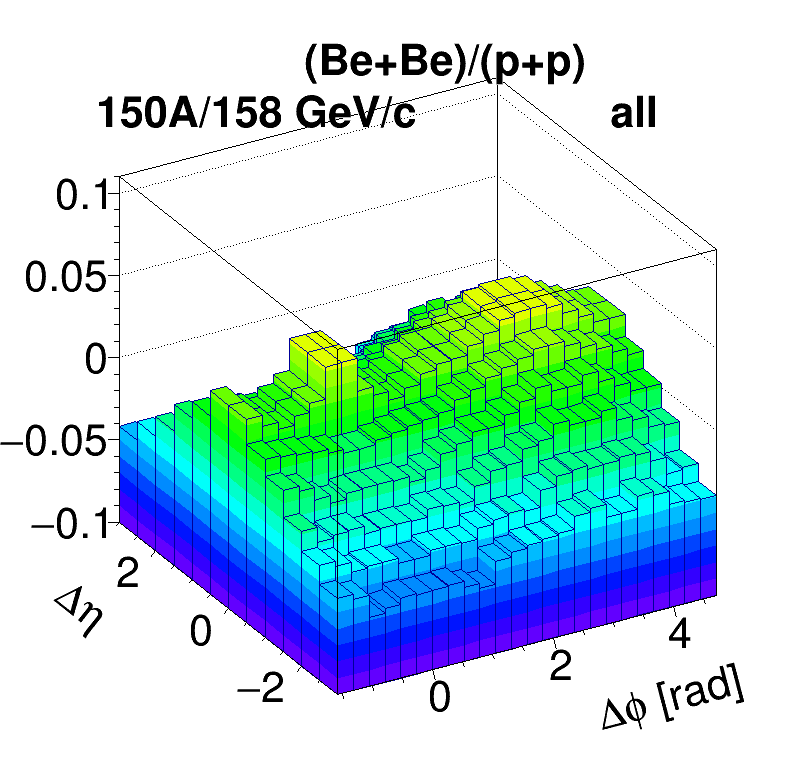}
  \includegraphics[width=0.24\textwidth]{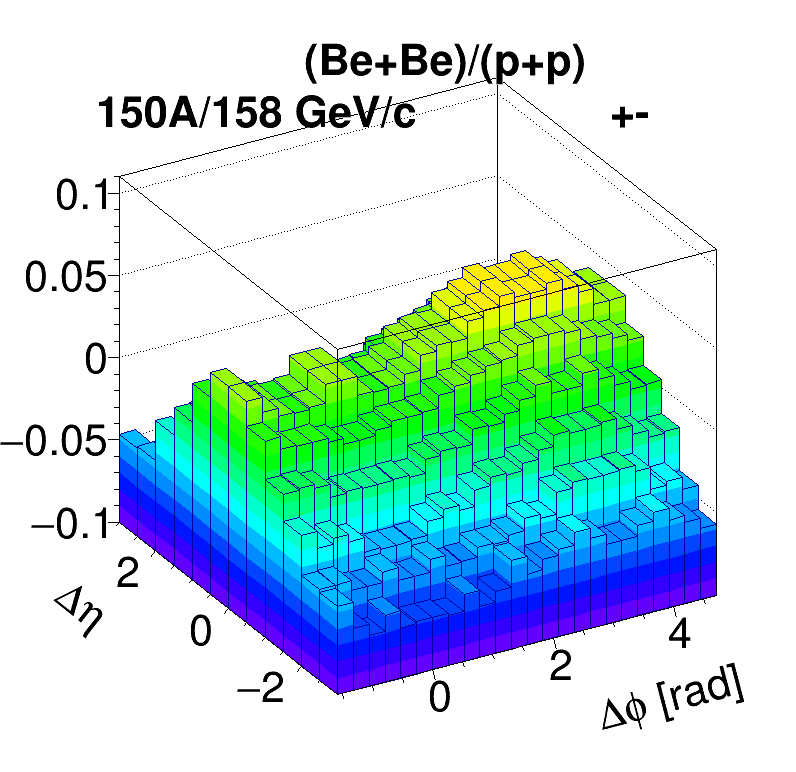}
  \includegraphics[width=0.24\textwidth]{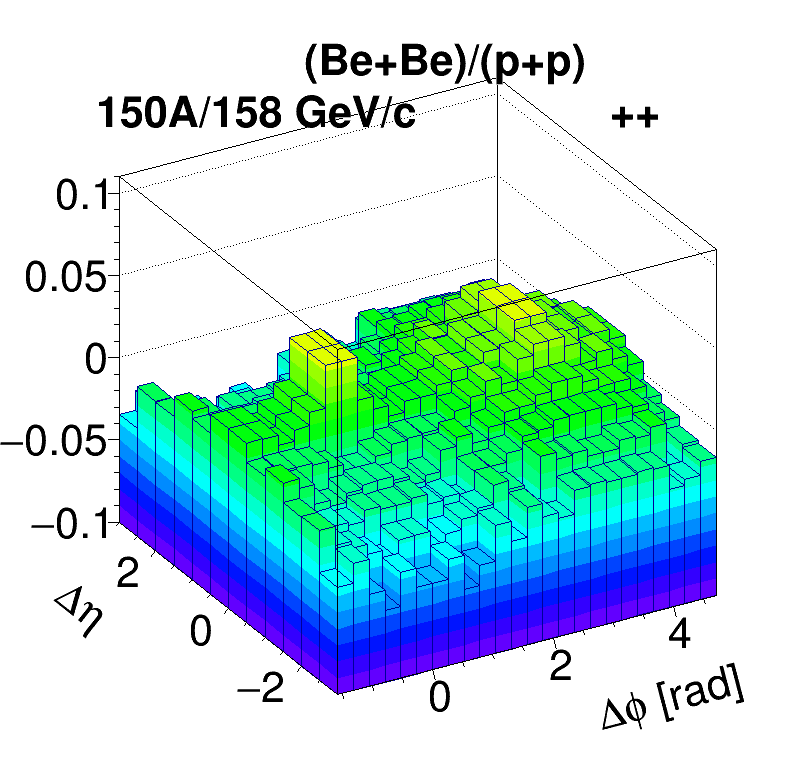}
  \includegraphics[width=0.24\textwidth]{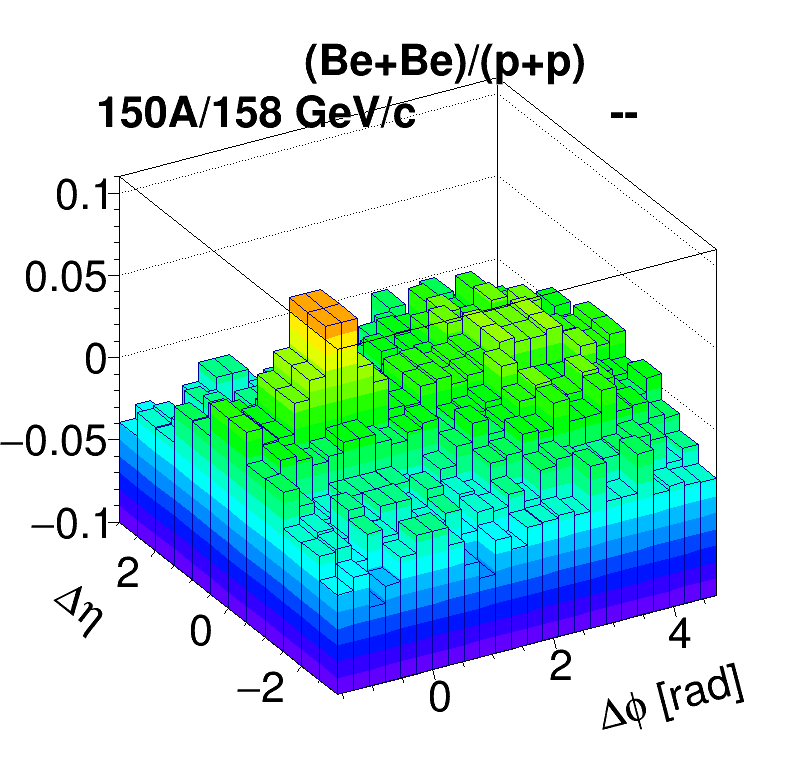}

  \caption{(Color online) Ratio $R^{\text{BeBe}}_{\text{pp}}$ for all pair combinations and momenta (following formula Eq.~\ref{eq:ratio_BeBe_pp}). Each row shows different beam momentum (beam momentum increases downwards). Every column presents results for different pair combinations (from left to right: all pairs, unlike-sign pairs, positive pairs, negative pairs). Note that correlation function for p+p was scaled down by a factor of 5 before calculation of ratio.} 
  \label{fig:ratio_pp_BeBe}
\end{figure*}

The results for all pair combinations and beam momenta are shown in Fig.~\ref{fig:ratio_pp_BeBe}. The near-side peak structure clearly increases in height with beam momentum, demonstrating that the contribution to that region in Be+Be is stronger than in p+p reactions.

\begin{figure*}
  \centering
  \includegraphics[trim=1.5cm 3cm 0cm 1cm, width=\textwidth]{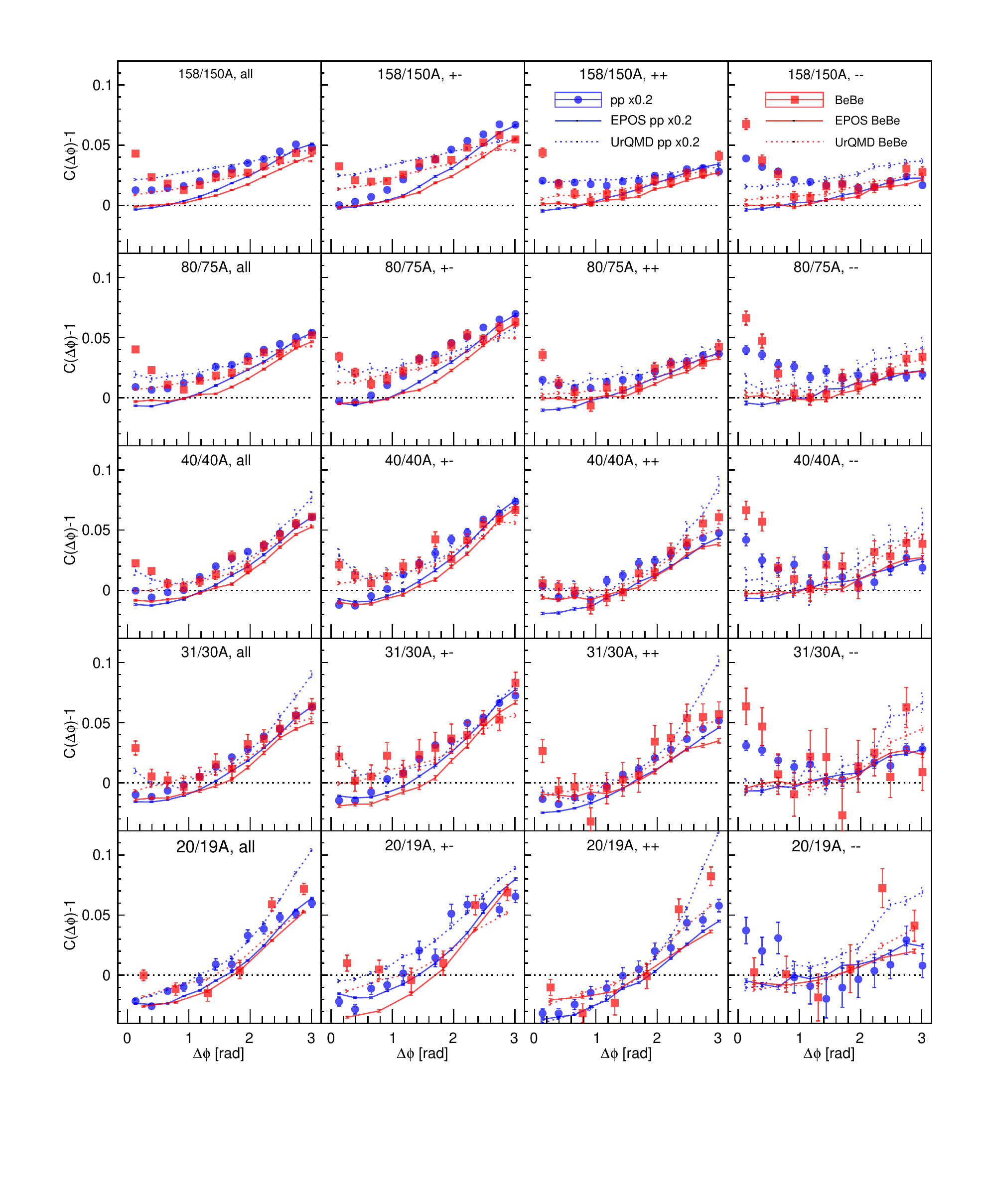}
  \caption{(Color online) Comparison of two-particle correlation function $C-1$ for the range $\Delta\eta \in [0, 0.5)$ in collisions of p+p (blue) and Be+Be (red). Columns from left to right show results for: all charge pairs, unlike-sign pairs, positive pairs, and negative pairs. Results for different beam momenta are plotted in successive rows. Data results are shown by markers (circles for p+p and squares for Be+Be), model results by lines (solid lines show \Epos while dotted lines -- UrQMD results). The results for p+p interactions were additionally scaled down by a factor of 5. Only statistical uncertainties are shown.}
  \label{fig:BeBe_pp_first_bin_comparison}
\end{figure*}

The correlation functions $C-1$ for both colliding systems is presented in Fig.~\ref{fig:BeBe_pp_first_bin_comparison} in a near-side slice of $0 \leq \Delta\eta \leq 0.5$ versus $\Delta\phi$. The values of $C$ for p+p interactions were scaled down by a factor of 5 to approximately account for the expected dilution effect.

In Sec.~\ref{sec:detadphi_results} the HBT+Coulomb+FS effects were pointed out as 
a possible source of the peak in the near-side region.
These correlations are indeed of importance for small relative four momentum and, as a consequence, 
for small relative transverse momentum $\Delta p_T$ of the hadron pair. They can produce a maximum 
in $\Delta p_T$ at about 20~\MeVc, stretching out to about 50~\MeVc 
(see e.g. Refs.~\cite{Kincses:2019rug,Lisa:2005dd,Martin:1998kk,Sinyukov:1998fc}).
 
The relative azimuthal angle is related to the relative transverse momenta of
the hadron pair via the relation:
\begin{equation}
\Delta p_{T}^2 = p_{T1}^{2} + p_{T2} ^{2} - 2p_{T1} p_{T2} \cos(\Delta\phi)
\end{equation}
\begin{equation}
\Delta\phi = \arccos \left(1 - \frac{\Delta p_T^2}{2 p_{T}^2} \right)
\end{equation}
For $\Delta p_{T}= 0.05$~\GeVc and the most probable value of transverse momenta of about 0.4~\GeVc
this corresponds to a difference in azimuthal angle $\Delta\phi$ of about 0.1~rad.
Therefore HBT+Coulomb+FS correlations can explain only a fraction of the enhancement in the 
first $\Delta\phi$ bin in the panels of Fig.~\ref{fig:BeBe_pp_first_bin_comparison}.
A similar conclusion was drawn e.g. in Refs.~\cite{Adams:2006tj,Acharya:2018ddg} at RHIC 
and LHC energies in nucleus-nucleus and p+A interactions. There a cut on $\Delta\phi$ was applied 
to remove HBT+C+FS effects in order to allow cleaner studies of processes like mini-jet production.

In the Fig.~\ref{fig:BeBe_pp_first_bin_comparison} we observe larger near-side correlations for negative charge than for positive charge pairs.
However, one should remember that the total number of negative charge pairs is about 2 to 3 times smaller than that of the positive charge pairs.
This is expected,  because of the proton admixture in the positive charge pair sample and, in consequence, contributions of $pp$ and $p\pi^+$ pairs in addition to $\pi^+\pi^+$ combinations.
It is not clear, however, why the contribution to the correlation peak from negative charge pairs
($\Delta\phi < 7.5^\circ$) seems to be smaller ($ \approx 10 \pm 2 \%$) than that from positive charge pairs ($ \approx 20\pm 2 \%$).
We hope to understand better the observed difference with the larger statistics data from Ar+Sc collisions.

\begin{figure*}
  \centering
  \includegraphics[trim=1.5cm 0cm 0cm 0cm, width=\textwidth]{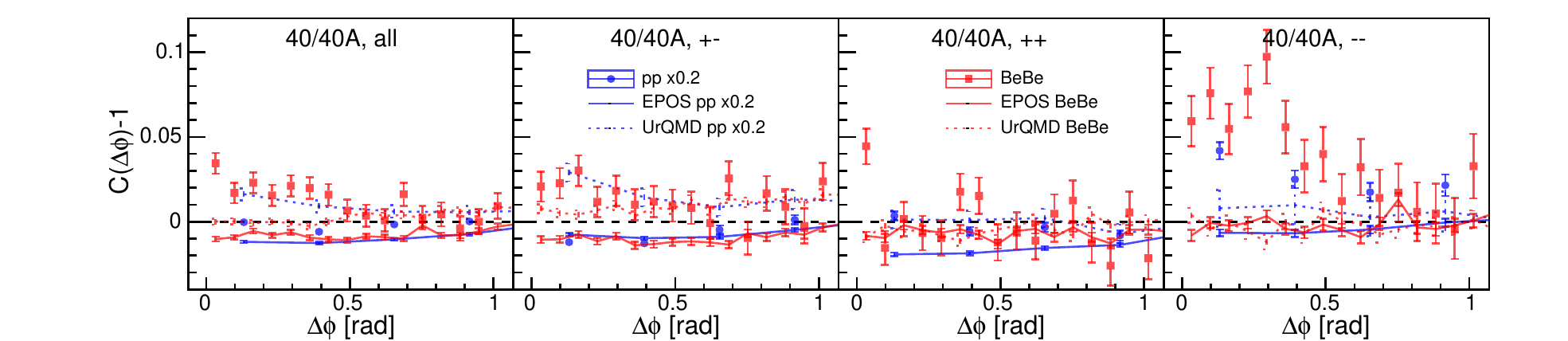}\\
  \includegraphics[trim=1.5cm 0cm 0cm 0cm, width=\textwidth]{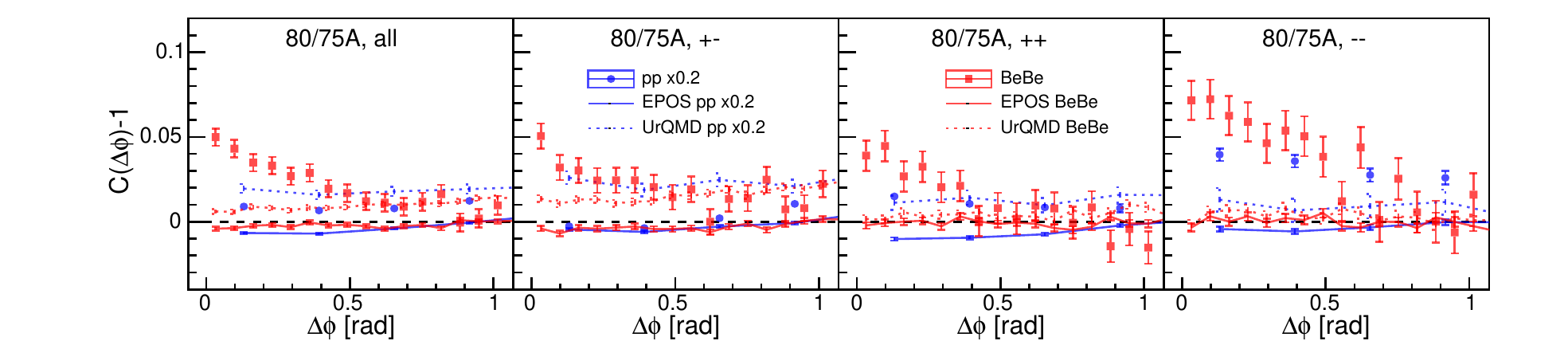}\\
  \includegraphics[trim=1.5cm 0cm 0cm 0cm, width=\textwidth]{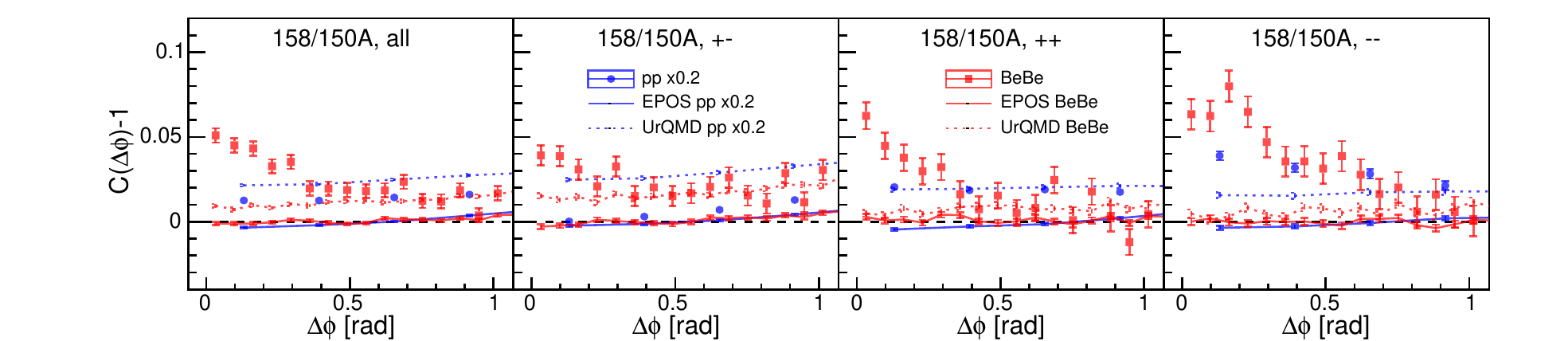}
  \caption{(Color online) Comparison of two-particle correlation function $C-1$ for the range $\Delta\eta \in [0, 0.5)$ and for zoomed in range of $0 \leq \Delta\phi \leq 1$ radians. Results of collisions of p+p are marked in blue and of Be+Be are marked in red. Columns from left to right show results for: all charge pairs, unlike-sign pairs, positive pairs, and negative pairs. Results for beam momenta 40$A$, 75$A$, and 150\AGeVc are plotted in successive rows. Data results are shown by markers (circles for p+p and squares for Be+Be), model results by lines (solid lines show \Epos while dotted lines -- UrQMD results). The results for p+p interactions were additionally scaled down by a factor of 5. Only statistical uncertainties are shown.}
  \label{fig:BeBe_pp_first_bins_zoomed}
\end{figure*}

Due to limited statistics, in particular for 19$A$ and 30\AGeVc, rather large bins in $\Delta\phi$ were used in our analysis.
To show better the influence of the HBT+Coulomb+FS mechanisms the results of Fig.~\ref{fig:BeBe_pp_first_bin_comparison} are shown in Fig.~\ref{fig:BeBe_pp_first_bins_zoomed} with finer binning (the width of $\Delta\phi$ bin here is $3.75^{\circ} \approx 0.07$ rad).
This figure presents the region of $\Delta\phi$ up to 1 radian for beam momenta of 40$A$, 75$A$, and 150\AGeVc.
One finds that the fraction of the two first bins corresponds to $(10.5 \pm 1.5)\%$, $(15.6\pm 2.0)\%$ and $(16.6\pm2.0)\%$ of the whole near-side effect for 40$A$, 75$A$, and 150\AGeVc respectively, for the sum of all charge configurations. This excludes the dominant role of the HBT+Coulomb+FS mechanisms in the near-side region.

\begin{figure*}
  \centering
  \includegraphics[width=0.6\textwidth]{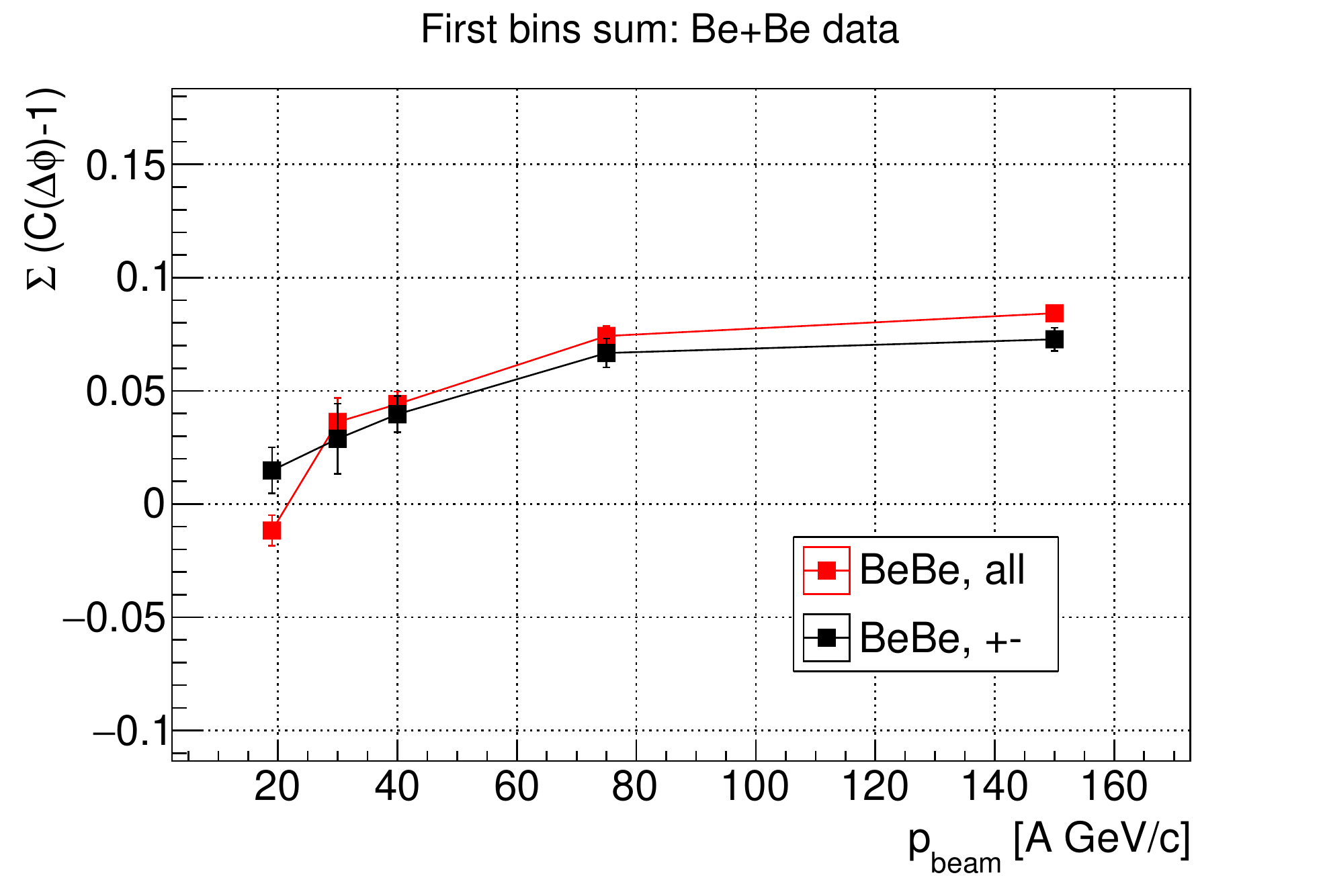}
  \caption{(Color online) The sum of two-particle correlation function $C(\Delta\eta,\Delta\phi)$ in 
the $\Delta\phi$ region $[0,\pi/2)$ and $\Delta\eta \in [0,0.5)$ for unlike-charge pairs (black squares) 
and all charge pairs (red circles) for the 5\% most central Be+Be collisions.}
  \label{fig:Sum}
\end{figure*}

To better visualise the energy dependence of the near-side correlations the first three points 
from the Fig.~\ref{fig:BeBe_pp_first_bin_comparison} were summed up for different charge configurations 
of the hadron pairs. The result is shown in Fig.~\ref{fig:Sum} for all charge and unlike-sign pairs. 
A strong energy dependence is observed between 19$A$-30$A$-40\AGeVc and 75$A$-150\AGeVc which could favour 
the onset of mini-jet formation. If the mini-jet hypothesis is correct, the correlation for oppositely 
charged particles may be due to local charge conservation.

\begin{figure*}
  \centering
  \includegraphics[width=0.49\textwidth]{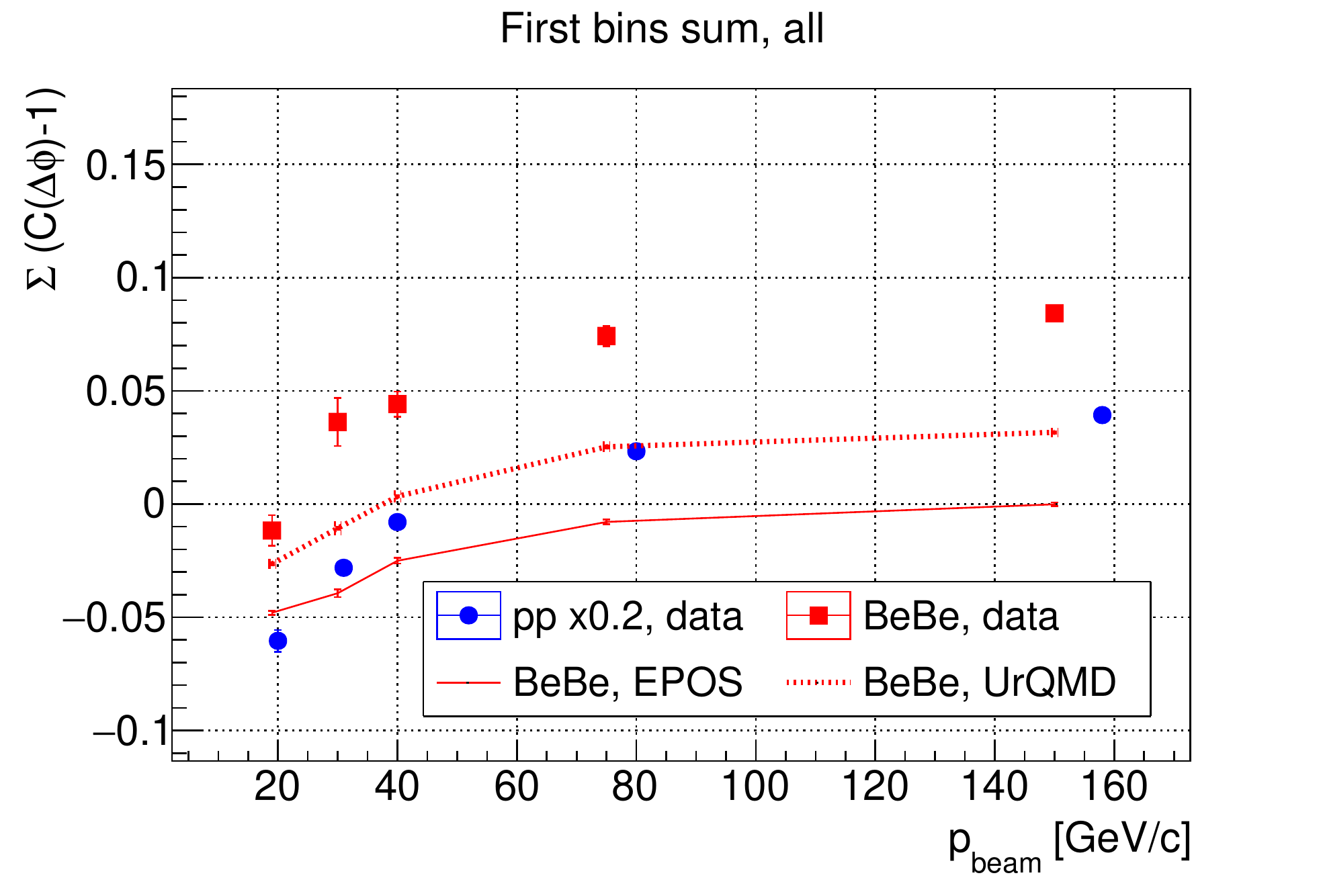}
  \includegraphics[width=0.49\textwidth]{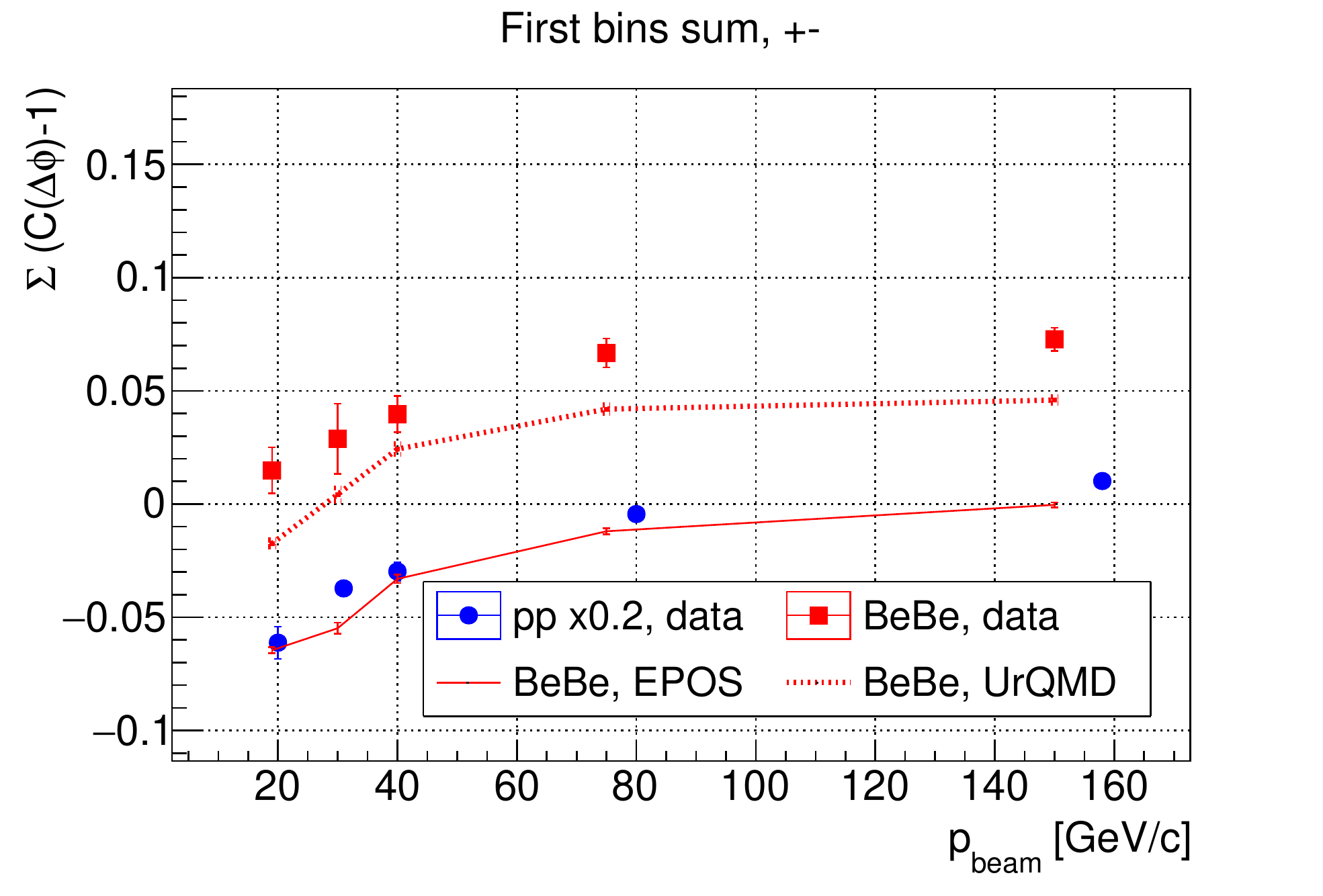}
  \includegraphics[width=0.49\textwidth]{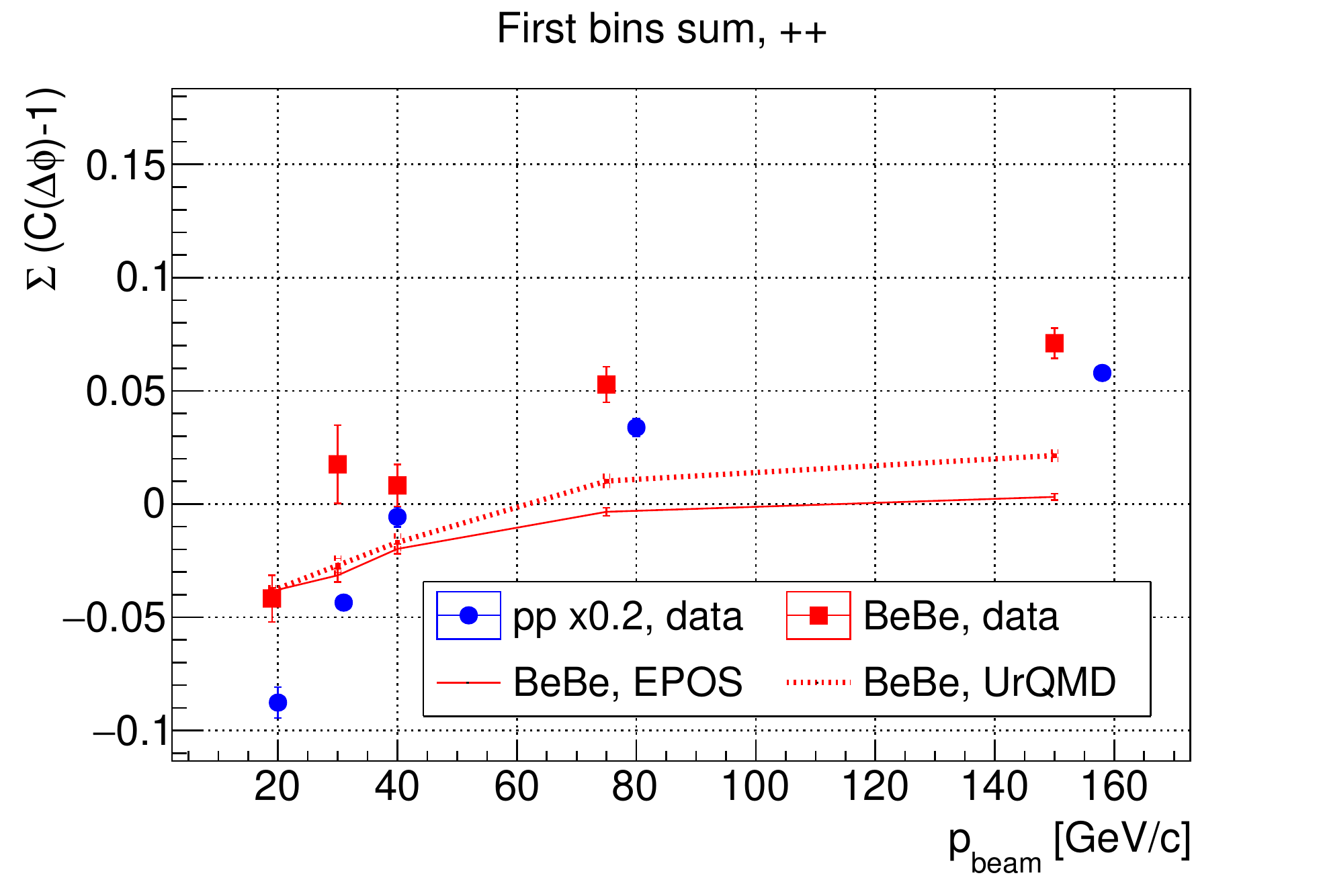}
  \includegraphics[width=0.49\textwidth]{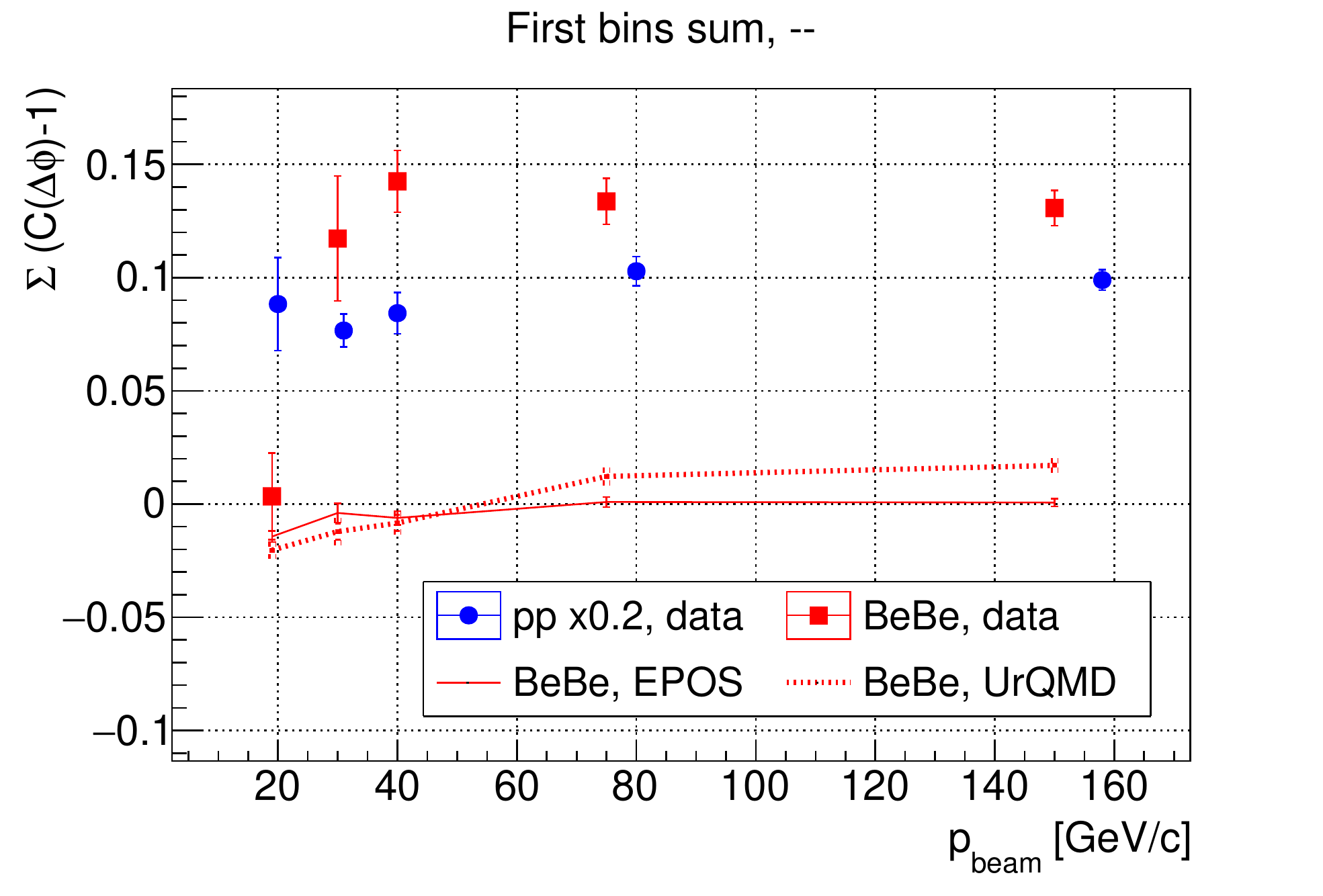}
  \caption{(Color online) The sum of two-particle correlation function $C(\Delta\eta,\Delta\phi)$ in the $\Delta\phi$ 
region $[0,\pi/2)$ and $\Delta\eta \in [0,0.5)$ for different charge combinations. Top left picture 
for all charge pairs, top right for unlike-sign, bottom left for positive charged pairs, 
bottom right for negative charged. Results for Be+Be (red squares) were compared with p+p (blue circles) 
and with model predictions of the corresponding charge combination. The results of p+p are scaled down by factor of 5.}
  \label{fig:Sum_with_pp}
\end{figure*}

In Fig.~\ref{fig:Sum_with_pp} the energy dependence of that sum of $C-1$ in Be+Be collisions for the data 
and models is presented for different charge configurations. In the same figure the experimental results 
for proton-proton collisions divided by a factor of 5 (i.e. $C/5-1$) are plotted.

\subsection{Comparison with models}

Since hard-scattering processes are not expected to contribute substantially to particle
production at SPS energies comparison of data with models designed to work also at low
collision energies, such as \Epos and UrQMD, appear to be most appropriate. One needs to point
out that unfortunately the models do not incorporate HBT+Coulomb+FS effects. Correlations in UrQMD and \Epos
are probably dominated by resonance production and string fragmentation processes. 
The resonance contribution was taken into account in both UrQMD and \Epos models, but discrepancies 
between data and simulations persist (see discussion below).
 
Although SPS energies are rather low, other models, which take into account the possibility of 
quark-gluon phase formation, were also proposed to explain the observed near-side correlations. 
These were interpreted as a result of 
jet fragmentation or color-tube fragmentation~\cite{Adams:2005dq,Marquet:2011gn}, 
originating from a Glasma flux tube~\cite{Dumitru:2008wn} or from 
Parton Bubbles~\cite{Lindenbaum:2000uq,Lindenbaum:2003ma,Lindenbaum:2008sx}. 
The latter were initially proposed by Van Hove \cite{VanHove:1984zy}.

\subsection{Comparison with the \Epos and UrQMD models}

\begin{figure*}
  \centering
  \includegraphics[trim=1.5cm 3cm 0cm 1cm, width=\textwidth]{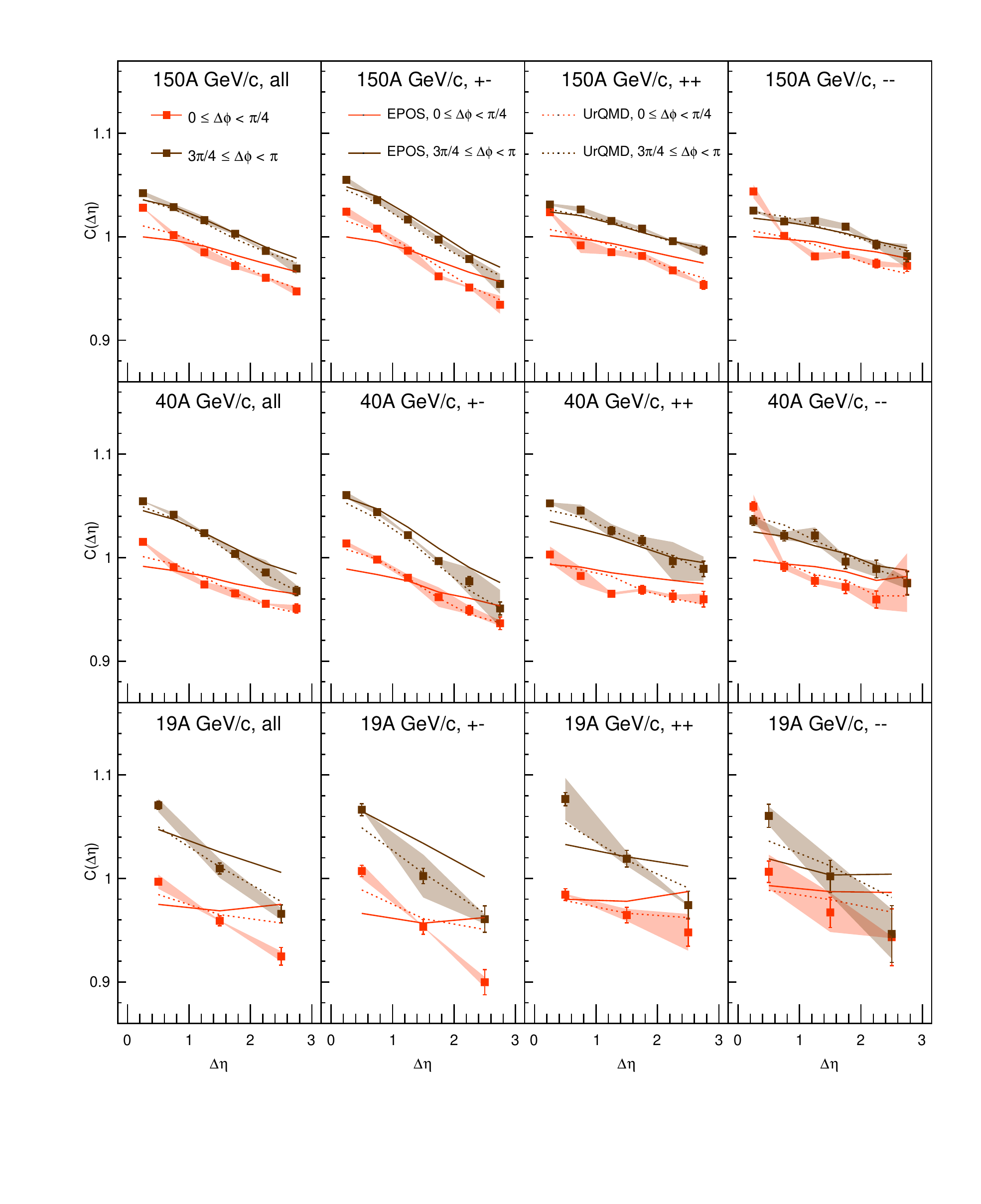}
  \caption{(Color online) Two-particle correlation function $C(\Delta\eta)$ obtained from projection of $C(\Delta\eta,\Delta\phi)$ onto the $\Delta\eta$ axis for subranges of $\Delta\phi$. From left to right the columns show respectively: all charge pairs, unlike-sign pairs, positive charge pairs, and negative charge pairs. Vertical bars denote statistical and shaded regions denote systematic uncertainties. Predictions of the \Epos model are shown by solid curves and the UrQMD model by dotted curves. Legend applies to all panels.}
  \label{fig:deta_projections}
\end{figure*}

\begin{figure*}
  \centering
  \includegraphics[trim=1.5cm 3cm 0cm 1cm, width=\textwidth]{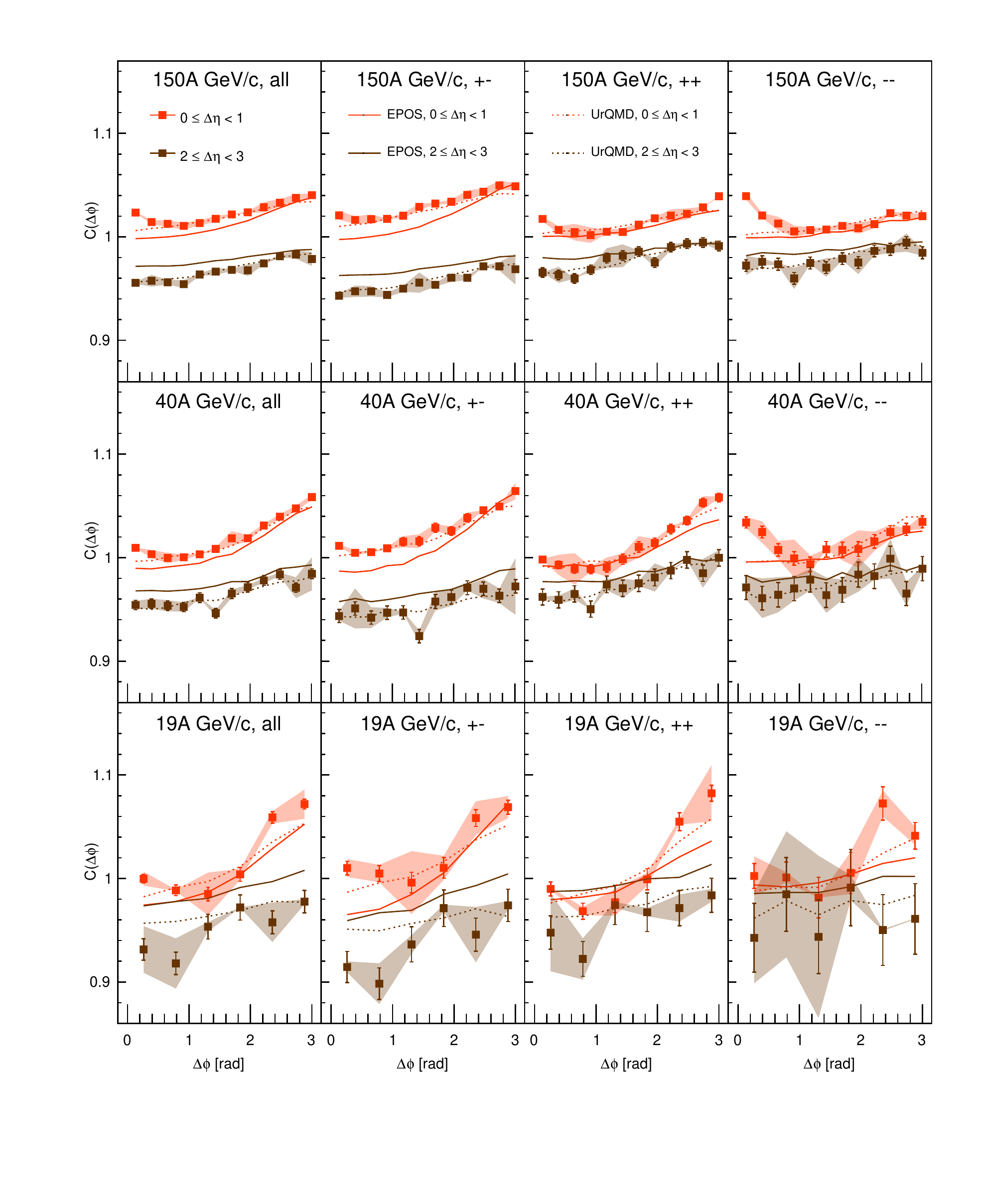}
  \caption{(Color online) Two-particle correlation function $C(\Delta\phi)$ obtained from projection of $C(\Delta\eta,\Delta\phi)$ onto the $\Delta\phi$ axis for subranges of $\Delta\eta$. From left to right the columns show respectively: all charge pairs, unlike-sign pairs, positive charge pairs, and negative charge pairs. Vertical bars denote statistical and shaded regions denote systematic uncertainties. Predictions of the \Epos model are shown by solid curves and the UrQMD model by dotted curves. Legend applies to all panels.}
  \label{fig:dphi_projections}
\end{figure*}

The final corrected results on $C(\Delta\eta,\Delta\phi)$ were compared with predictions of 
the pure (i.e.~not reconstructed) \Epos and UrQMD models. They are presented in the form of projections 
on the $\Delta\eta$ and $\Delta\phi$ axes in Fig.~\ref{fig:deta_projections} and \ref{fig:dphi_projections}, 
respectively. In case of projection onto the $\Delta\eta$ axis the correlations were divided 
into four sub-ranges of $\Delta\phi$: $0 \leq \Delta\phi < \frac{\pi}{4}$, 
$\frac{\pi}{4} \leq \Delta\phi < \frac{\pi}{2}$, $\frac{\pi}{2} \leq \Delta\phi < \frac{3\pi}{4}$, 
and $\frac{3\pi}{4} \leq \Delta\phi < \pi$. In case of projection onto the $\Delta\phi$ axis, 
the $\Delta\eta$ axis was divided into three sub-ranges: $0 \leq \Delta\eta < 1$, 
$1 \leq \Delta\eta < 2$, and $2 \leq \Delta\eta < 3$. The functions

\begin{equation}
  \label{eq:deta}
  C(\Delta\eta)=
  \frac{N_{\text{mixed}}^{\text{pairs}}}{N_{\text{data}}^{\text{pairs}}}
  \frac{D(\Delta\eta)}{M(\Delta\eta)}
\end{equation}

 and

\begin{equation}
  \label{eq:dphi}
  C(\Delta\phi)=
  \frac{N_{\text{mixed}}^{\text{pairs}}}{N_{\text{data}}^{\text{pairs}}}
  \frac{D(\Delta\phi)}{M(\Delta\phi)}
\end{equation}

were recalculated in those sub-ranges. Full statistical uncertainty analysis was performed in each sub-range as well. In case of systematic uncertainty analysis, the general method of calculation was similar to the one mentioned in Sec.~\ref{sec:systematic}, however here the mean systematic uncertainty was not calculated. The difference of ($\text{loose} - \text{tight}$) was shown in each bin separately instead. In order to present the results in a clear way, only two sub-ranges 
are shown in Figs.~\ref{fig:deta_projections} and~\ref{fig:dphi_projections}.

 The UrQMD model predictions are in reasonable agreement with the measured data for both 40$A$ and 150\AGeVc 
(dashed lines in Figs.~\ref{fig:deta_projections} and \ref{fig:dphi_projections}) for all charge combinations. 
However some differences are seen in the region of the smallest $\Delta\phi$ and $\Delta\eta$ points.
The \Epos model works well for away-side correlations (large $\Delta\eta$ and $\Delta\phi$) and fails 
in the near-side region (full lines in Figs.~\ref{fig:deta_projections} and \ref{fig:dphi_projections}).
At the smallest beam momentum 19\AGeVc, the deviations between data and both models are quite large, 
but the statistical uncertainties are large and the models are not intended for this low energy.
Such a discrepancy between both models predictions and the data exists as well for the 
region $\Delta\eta < 0.5$, $\Delta\phi < \pi/2$ as evident from Fig.\ref{fig:BeBe_pp_first_bin_comparison}. 
It appears for both Be+Be and p+p interactions and different charge combinations.

The comparison between model predictions and the Be+Be data in the region of small $\Delta\eta$ 
and $\Delta\phi < \pi/2$ is shown in Fig.~\ref{fig:Sum_with_pp}. Both models underpredict 
the correlation strength for all the energies measured, but the discrepancy is significantly 
larger for \Epos predictions.

\section{Summary}\label{sec:summary}

The $C(\Delta\eta,\Delta\phi)$ correlation function was studied by the \NASixtyOne experiment in the 0-5\% most central Be+Be collisions for a range of different incident beam momenta: 19$A$, 30$A$, 40$A$, 75$A$, and 150\AGeV\c.
Near-side and away-side correlations were observed and measured as a function of incident momentum and particle charge combinations.
The strength of the correlation was compared with that observed by \NASixtyOne in p+p~\cite{Aduszkiewicz:2016mww} interactions for which the same experimental conditions were applied.
A strong suppression of the correlation function is observed as compared with p+p interaction.

In Be+Be interactions, one observes a back-to-back correlation which is rather wide in $\Delta\phi$ 
and decreases with energy (see Fig.~\ref{fig:deta_projections}, \ref{fig:dphi_projections}). 
The correlation function behaviour of the data is qualitatively described 
by the UrQMD 3.4 and \Epos 1.99 models. The UrQMD model shows better quantitative agreement.

A narrow enhancement around $(\Delta\eta,\Delta\phi)=(0,0)$ appears clearly 
for Be+Be collisions (see Figs.~\ref{fig:dphi_projections}, \ref{fig:BeBe_pp_first_bin_comparison}). 
It is wider (RMS about $25^{\circ} \approx 0.45$~rad) in $\Delta\phi$ than expected 
from the contribution of HBT+Coulomb+FS correlations alone and increases with incident energy (see Fig.~\ref{fig:Sum}).
This observation disagrees with predictions of the studied models, which do not include quantum statistics and Coulomb repulsion effects, but simulate only the production of the majority of hadron resonances.
Much stronger near-side correlations were observed at higher energies 
 in nucleus-nucleus and p+A interactions~\cite{Adams:2006tj,Acharya:2018ddg}.
The \NASixtyOne experiment continues correlation studies in $\Delta\eta,\Delta\phi$
for Ar+Sc and Xe+La collisions. The results for different nucleus sizes 
may help to obtain more insight into the importance of the proposed mechanisms.

\section*{Acknowledgements}
We would like to thank the CERN EP, BE, HSE and EN Departments for the
strong support of NA61/SHINE.

This work was supported by
the Hungarian Scientific Research Fund (grant NKFIH 123842\slash123959),
the Polish Ministry of Science
and Higher Education (grants 667\slash N-CERN\slash2010\slash0,
NN\,202\,48\,4339 and NN\,202\,23\,1837), the National Science Centre Poland (grants~2014\slash14\slash E\slash ST2\slash00018, 2014\slash15\slash B\slash ST2 \slash\- 02537 and
2015\slash18\slash M\slash ST2\slash00125, 2015\slash 19\slash N\slash ST2\slash01689, 2016\slash23\slash B\slash ST2\slash00692,
2017\slash\- 25\slash N\slash\- ST2\slash\- 02575,
2018\slash 30\slash A\slash ST2\slash 00226,
2018\slash 31\slash G\slash ST2\slash 03910),
the Russian Science Foundation, grant 16-12-10176 and 17-72-20045,
the Russian Academy of Science and the
Russian Foundation for Basic Research (grants 08-02-00018, 09-02-00664
and 12-02-91503-CERN),
the Russian Foundation for Basic Research (RFBR) funding within the research project no. 18-02-40086,
the National Research Nuclear University MEPhI in the framework of the Russian Academic Excellence Project (contract No.\ 02.a03.21.0005, 27.08.2013),
the Ministry of Science and Higher Education of the Russian Federation, Project "Fundamental properties of elementary particles and cosmology" No 0723-2020-0041,
the European Union's Horizon 2020 research and innovation programme under grant agreement No. 871072,
the Ministry of Education, Culture, Sports,
Science and Tech\-no\-lo\-gy, Japan, Grant-in-Aid for Sci\-en\-ti\-fic
Research (grants 18071005, 19034011, 19740162, 20740160 and 20039012),
the German Research Foundation (grant GA\,1480/8-1), the
Bulgarian Nuclear Regulatory Agency and the Joint Institute for
Nuclear Research, Dubna (bilateral contract No. 4799-1-18\slash 20),
Bulgarian National Science Fund (grant DN08/11), Ministry of Education
and Science of the Republic of Serbia (grant OI171002), Swiss
Nationalfonds Foundation (grant 200020\-117913/1), ETH Research Grant
TH-01\,07-3 and the Fermi National Accelerator Laboratory (Fermilab), a U.S. Department of Energy, Office of Science, HEP User Facility managed by Fermi Research Alliance, LLC (FRA), acting under Contract No. DE-AC02-07CH11359 and the IN2P3-CNRS (France).


\providecommand{\href}[2]{#2}\begingroup\raggedright\endgroup

\newpage
{\Large The \NASixtyOne Collaboration}
\bigskip
\begin{sloppypar}

\noindent
A.~Aduszkiewicz$^{\,15}$,
E.V.~Andronov$^{\,21}$,
T.~Anti\'ci\'c$^{\,3}$,
V.~Babkin$^{\,19}$,
M.~Baszczyk$^{\,13}$,
S.~Bhosale$^{\,10}$,
A.~Blondel$^{\,4}$,
M.~Bogomilov$^{\,2}$,
A.~Brandin$^{\,20}$,
A.~Bravar$^{\,23}$
% ...

\end{sloppypar}

\noindent
% ...
$^{2}$~Faculty of Physics, University of Sofia, Sofia, Bulgaria\\
$^{3}$~Ru{\dj}er Bo\v{s}kovi\'c Institute, Zagreb, Croatia\\
$^{4}$~LPNHE, University of Paris VI and VII, Paris, France\\
% ...
$^{10}$~Institute of Nuclear Physics, Polish Academy of Sciences, Cracow, Poland\\
% ...
$^{13}$~AGH - University of Science and Technology, Cracow, Poland\\
% ...
$^{15}$~University of Warsaw, Warsaw, Poland\\
% ...
$^{19}$~Joint Institute for Nuclear Research, Dubna, Russia\\
$^{20}$~National Research Nuclear University (Moscow Engineering Physics Institute), Moscow, Russia\\
$^{21}$~St. Petersburg State University, St. Petersburg, Russia\\
% ...
$^{23}$~University of Geneva, Geneva, Switzerland\\
% ...

\end{document}